\documentclass[aps,prx,superscriptaddress, twocolumn
]{revtex4-2}
\usepackage{graphicx, xcolor, hyperref}
\usepackage{amsmath,amssymb,MnSymbol,bbold}
\usepackage{dsfont}
\usepackage{braket}
\usepackage{enumerate}

\newcommand{\figref}[1]{Fig.~\ref{#1}}
\newcommand{\secref}[1]{Sec.~\ref{#1}}
\newcommand{\tabref}[1]{Table \ref{#1}}
\renewcommand{\eqref}[1]{Eq.~(\ref{#1})}

\newcommand{\ip}{\mathrm{I}_\mathrm{p}}
\newcommand{\up}{\mathrm{U}_\mathrm{p}}
\newcommand{\pb}{\mathbf{p}}

\newcommand{\dd}{\mathrm{d}}
\newcommand{\rb}{\mathbf{r}}

\newcommand{\Ab}{\mathbf{A}}

\renewcommand\Re{\mathrm{Re}}

\newcommand{\erf}{\;\mathrm{erf}}
\newcommand{\id}{\mathds{1}}

\newcommand*\conj[1]{\overline{#1}}
\begin{document}
	
	\title{Quantum estimation in strong fields: \textit{in situ} ponderomotive sensing}
	
	\author{A. S. Maxwell}
	\email{andrew.maxwell@ucl.ac.uk}
	\affiliation{Institut de Ciencies Fotoniques, The Barcelona Institute of Science and Technology, 08860 Castelldefels (Barcelona), Spain}
	\affiliation{Department of Physics \& Astronomy, University College London, Gower Street, London, WC1E 6BT, United Kingdom}

	\author{A. Serafini}
	\affiliation{Department of Physics \& Astronomy, University College London, Gower Street, London, WC1E 6BT, United Kingdom}
	
	\author{S. Bose}
	\affiliation{Department of Physics \& Astronomy, University College London, Gower Street, London, WC1E 6BT, United Kingdom}
	
	\author{C. Figueira de Morisson Faria}
	\affiliation{Department of Physics \& Astronomy, University College London, Gower Street, London, WC1E 6BT, United Kingdom}
	
	\date{\today}
	
	\begin{abstract}
		We develop a new theoretical framework to optimize and understand uncertainty from \textit{in situ} strong-field measurements of laser field parameters. We present the first derivation of quantum and classical Fisher information in attoscience for an electron undergoing strong-field ionization.
		This is used for parameter estimation and to characterize the uncertainty of the ponderomotive energy, directly proportional to laser intensity. 
		In particular, the quantum and classical Fisher information for the momentum basis displays quadratic scaling over time.		
		This can be linked to above-threshold ionization interference rings for measurements in the momentum basis and to a `ponderomotive phase' for the optimal quantum measurements. Preferential scaling in uncertainty is found for increasing laser pulse length and intensity. 
		We use this to demonstrate for \textit{in situ} measurements of laser intensity, that high resolution momentum spectroscopy has the capacity to reduce the uncertainty by over 25 times compared to measurements employing the ionization rate, while using the optimal quantum measurement would reduce it by a further factor of 2.6.
		A minimum uncertainty of the order $2.8 \times 10^{-3}$\% is theorized for this framework.
		Finally, we examine previous \textit{in situ} measurements formulating a measurement that matches the experimental procedure and suggest alterations to the measurement scheme that could reduce the laser intensity uncertainty.
	\end{abstract}

	\maketitle

\section{Introduction}
The time-averaged quiver energy of an electron in an external laser field, known as the ponderomotive energy $U_p$, is a key concept in strong-field laser-matter interaction (intensities in the range $10^{13} - 10^{20}$~W/cm$^2$). For instance, cutoff laws for  phenomena such as above-threshold ionization (ATI) \cite{MilosReviewATI,Becker2002Review,Becker2018} and high-order harmonic generation (HHG) \cite{Lewenstein1994}, and momentum constraints in nonsequential double and multiple ionization (NSDI, NSMI) \cite{Shaaran2010,Shaaran2010a,Faria2011} are universal functions of $U_p$. This stems from the physical mechanism behind them, namely the laser-induced recollision or recombination of an electron with its parent ion \cite{Corkum1993,Becker1995}. The active electron is freed close to the peak of the field, propagates in the continuum and it is driven back to the target near a field crossing. The kinetic energy acquired in the continuum then is released upon a non-linear interaction with the target, either as a high-energy photon or photoelectron. Because a typical field cycle is roughly a few femtoseconds, these processes occur within hundreds of attoseconds ($10^{-18}$~s).
 
  Attoseconds are roughly the time it takes for an electron to travel through atomic distances. Thus, strong-field phenomena allow an unprecedented control over real-time electron dynamics  \cite{Lein2007,Krausz2009,Salieres2012R}. They provide tools for dealing with fundamental questions such as tunneling times \cite{Ni2018,Hofmann2019,Sainadh2019}, or electron migration in small \cite{Smirnova2009,Mairesse2010} and larger molecules \cite{Calegari2014,Lepine2014,Kuleff2016,Calegari2016Migration}, and may pave the way for optoelectronic computers up to 10,000 times faster than existing devices  \cite{Schiffrin2013,Schultze2013}. Strong-field phenomena have also led to imaging applications such as ultrafast photoelectron holography \cite{HuismansScience2011,Faria2020} or high-harmonic spectroscopy \cite{Marangos2016}, which both display a high sensitivity to the target and the field.
  
  Unfortunately, many of the laser field parameters cannot be accurately measured. This is particularly true for the laser intensity \cite{Smeenk2011,Pullen2013}, where it is common for experiments to cite $10-20\%$ uncertainty. The ponderomotive energy is directly proportional to and usually determined by the laser intensity.  This lack of precision hinders progress, as the system's dynamics vary critically with $U_p$. A well-known example is the boundary between the multiphoton and tunneling regimes, determined by the Keldysh parameter $\gamma=\sqrt{I_p/(2U_p)}$, where  $I_p$ denotes the ionization potential, which indicates the electron's prevalent ionization mechanism (for a review see  \cite{Popruzhenko2014}). Other critical scenarios involve resonances with ponderomotively upshifted bound states  \cite{Faria2002,Milos2002,Wassaf2003,Taieb2003,Strelkov2010,Tudorovskaya2011,Strelkov2014}, or giant resonances in multielectron systems \cite{Shiner2011,Chen2015,Facciala2016,Chen2018}. The intensity also determines the below-threshold regime in  NSDI  \cite{Faria2011}, for which excitation and  quantum interference become important \cite{Hao2014,Maxwell2015,Maxwell2016}. Thus, it must be accurately known.	
  

  \textit{In situ} measurements are ideally suited to attoscience, as they exploit the high sensitivity and nonlinearity of strong-field phenomena with regard to the laser field intensity. While early work focused on the yields of singly and doubly charged ions \cite{Larochelle1998, Bhardwaj2001}, the photoelectron momentum distributions obtained from time of flight spectrometers, velocity map imaging or cold target recoil ion momentum spectroscopy (COLTRIMS) enabled momentum-based \textit{in situ} measurements. This made it possible to determine the driving-field intensity \cite{Litvinyuk2003} by mapping it to the final electron momentum via a classical equation of motion \cite{Corkum1989}. Including a quasi-static ionization rate and the temporal and spatial variation of the laser beam improved the accuracy and led to fitting errors of $10\%$ \cite{Alnaser2004}  and  $4\%$ \cite{Smeenk2011}, respectively. However, due to uncertainty in other experimental parameters an overall accuracy of $8\%$ was reported in \cite{Smeenk2011}. Important breakthroughs have been achieved for photoelectron spectra from atomic hydrogen employing a few-cycle linearly polarized pulse, where the experimental spectra were fitted to those computed with the time-dependent Schr\"odinger equation (TDSE) \cite{Pullen2013,Kielpinski2014,Khurmi2017}. This resulted in uncertainties of  $1\%$ in the intensity \cite{Pullen2013}, and of $2\%$ in the carrier-envelope phase \cite{Khurmi2017}. Recently, the first \textit{in situ} measurement of ellipticity was performed with an uncertainty of $0.7\%$ for xenon in a nearly circularly polarized field using ratios of intensity measurements \cite{Wang2019}. This was achieved by separately measuring the intensity of the two linear field components and taking the ratio.

Despite this progress, the majority of strong field experiments still report laser intensity uncertainties of around $10\%$. 
\textit{In situ} measurements can do better but the methods vary considerably depending on the experimental setup \cite{Kielpinski2014}. They often require ad-hoc relations or fitting processes, which (as black box approaches) give limited understanding of the mechanisms behind the laser intensity uncertainty.  
No general framework has been applied to date to identify and understand the main processes that affect such uncertainties and sensitivities in attoscience. Yet, such a framework does exist, and is provided by the methods of quantum metrology \cite{Giovannetti2011,Degen2017,Bongs2019}, which offer tools to evaluate and possibly improve the uncertainties. In mathematical terms, such methods are based on the evaluation of quantum and classical Fisher informations \cite{Helstrom1976,Paris2009}, yielding proportionality factors for the scaling of the uncertainty on a parameter’s estimation. 
Such methods have been applied to practical cases with great success, especially for optical fields in the quantum regime, leading to enhanced optical-phase tracking
\cite{Wheatley2010, Yonezawa1514} and, notably, to the super-resolution of incoherent light sources \cite{Tang:16,Paur:16,Yang:16,Tham2017,Donohue2018, Parniak2018}

Put simply, the classical Fisher information gives the uncertainty per measurement for a particular measurement set-up, while the quantum Fisher information yields the minimal possible uncertainty for any conceivable positive-operator valued measurement (POVM). These quantities can elucidate trends and physical mechanisms affecting the achieved uncertainty and may point to ways to reduce its value. It is important to emphasize that estimation schemes inspired by the evaluation of quantum and classical Fisher informations harness fully coherent quantum effects, and require full knowledge of the quantum state. Notably, quantum entanglement among distinct probes is well-known to improve the uncertainty scaling \cite{Giovannetti2011} but, less ambitiously, even coherence related to single-probe interference effects can be exploited to this aim. This is precisely what we set out to show in this study, by analysing the full quantum state within the validity range of the strong field approximation.
	
Quantum interference has \emph{indeed} proved a vital and integral part of strong field and attosecond physics.  Laser-induced ionization and/or recollision occurs at specific times within a laser cycle, which leads to well-defined phase shifts with regard to the field. Seminal examples are the high-order harmonic phase, which is critical for HHG propagation and attosecond-pulse generation \cite{Gaarde2008} and the inter-cycle interference rings that form in ATI photoelectron momentum distribution \cite{Lewenstein1995,Becker2002Review}. Furthermore, recollision at spatially separated centres leads to interference patterns that encode structural features in molecules \cite{Lein2007,Augstein2012}, and, recently, quantum interference has been reported in NSDI  \cite{Hao2014,Maxwell2015,Maxwell2016}. A key question within attosecond imaging of matter is how to retrieve quantum phase differences from photoelectron or high-harmonic spectra. Examples of interferometric schemes are the Reconstruction of Attosecond Burst By Interferences of Two-photons Transitions (RABBIT) technique \cite{Haessler2010}, the attosecond streak camera \cite{Itatani2002}, the Spectral Phase Interferometry for the Direct Electric Field Reconstruction (SPIDER) \cite{Cormier2005} and the Frequency Resolved  Optical  Gating (FROG) \cite{Gagnon2008}(for a review see \cite{Orfanos2019}), and time-resolved photoelectron holography, which uses phase differences between different electron pathways to extract information about the target \cite{HuismansScience2011,Hickstein2012}.
The strongest and most robust interference effects tend to be those directly related to the strong laser field. One primary example of such a quantum interference effect being ATI peaks/ rings. These are strongly dependent on the driving-field intensity \cite{Lewenstein1995} and can	 be used for intensity calibration. By computing the Fisher information using a quantum-orbit model we may turn this interference on or off and thus thoroughly investigate the effect on uncertainty in laser intensity measurements. 
	
In this article we will use the well-known workhorse of attoscience, the strong field approximation (SFA), see Ref.~\cite{Amini2019} for a review, as the underlying model with which to compute the quantum and classical Fisher information. To our knowledge the quantum and classical Fisher information have not been computed in the context of attoscience before. As such, in this first study, we will focus on the broader trends of the uncertainty due to the laser field rather than differences in targets, or quantitative comparisons with experiments. Instead, the present work aims at laying out a general theoretical framework to be employed in subsequent studies. A key point is that studies fitting experimental data to theoretical models to perform \textit{in situ} measurements, will  yield large uncertainties from inaccuracies in the model \cite{Kielpinski2014}. However, in the computation of quantum and classical Fisher information there is room for a qualitative analysis of the uncertainty, where the main trends and primary physical processes governing the uncertainty can be understood. The SFA is a model, which allows for this, in part because it is
approximate and neglects the effect of the binding potential on the photoelectrons. The simplicity of the SFA allows for analytical trends to be extracted and for a faster and easier construction of the Fisher information. 
The computation of the Fisher information requires the evaluation of high-oscillatory integrals, which leads to computing of the order of a million transition amplitudes per value calculated. Thus, it is not clear that it would be possible to use the numerical solutions of the TDSE for this purpose. Furthermore, the SFA allows switching quantum interference on and off at will. Coulomb-distorted orbit based methods are in principle applicable, but would provide additional challenges that lie outside the scope of the present work.
				
The structure of this article is as follows: In \secref{sec:Theory} the relevant theory for the SFA for a zero range potential is introduced (\secref{sec:Theory:StrongFieldApproximation}) and the quantum and classical Fisher information along with their relation to the uncertainty are defined (\secref{sec:Theory:FisherInformaition}). Next in \secref{sec:AnalyticalResults} the procedure for computing the quantum and classical Fisher information with the SFA (and similar orbit based models) is outlined. Additionally, using this definition some general trends are identified analytically. Following these analytical results in \secref{sec:NumericalResults} we present numerical calculations of the Fisher information. In \secref{sec:NumericalResults:GeneralTrends} we continue to explore the general trends and establish a strong link to  interference, while in \secref{sec:NumericalResults:Predicitons} we examine previous \textit{in situ} experiments to measure the laser intensity. Using the classical and quantum Fisher information we predict the minimum possible uncertainty and make suggestions on how the measurement could be improved by changing the experimental set-up, field parameters and the measurement basis. Finally, in \secref{sec:Conclusions} we sum up our findings, discuss their implication in the context of the recent strong field results and what the next step is with this framework.

	
\section{Background}
	\label{sec:Theory}
	
\subsection{Strong field approximation}
\label{sec:Theory:StrongFieldApproximation}

Throughout, we consider an electron in a strong laser field and a zero range potential, whose Hamiltonian (under the dipole approximation) reads
	\begin{equation}
	\hat{H}(t)=\frac{1}{2}\hat{\pb}^2+H_I(t)+\hat{V}
	\end{equation}
and  the zero range (regularized) potential is given by 
\begin{equation}
\hat{V}=-\frac{2\pi}{\sqrt{\ip}}\delta(\rb)\frac{\partial }{\partial r}r.
\end{equation}
The zero-range potential is advantageous for the implementation of the SFA, as it supports a single bound state and plane waves are a good approximation for the continuum; see \cite{Becker1994a} for more details. 
We employ atomic units throughout (denoted a.u.\ ), where the elementary charge, electron mass and $\hbar$ are set to one, $e=m=\hbar=1$.
The laser interaction Hamiltonian reads as  $H_I(t)=2\hat{\pb}\cdot\Ab(t)+\Ab(t)^2$ in the velocity gauge and $H_I(t)=\hat{\rb}\cdot\mathbf{E}(t)$ in the length gauge.
	The atomic Hamiltonian $H_0=\frac{1}{2}\hat{\pb}^2+\hat{V}$, describes the electron in the bound and scattering states. The Volkov Hamiltonian $H_V=\frac{1}{2}\hat{\pb}^2+H_I(t)$ describes a free electron propagating in a laser field \cite{Gordon1926,Volkov1935}.
	
	The electron starts in the only bound state of the zero range potential $\ket{\psi(t=0)}=\ket{0}$, with
	\begin{equation}
	\braket{r|0}=\frac{(\ip)^{1/4}\exp(-\sqrt{\ip}r)}{2\pi r}.
	\label{eq:ZeroPotBoundState}
	\end{equation}
	
	The full time evolution operator $U(t,0)$ can be written as
	\begin{equation}
	U(t,0)= U_{0}(t,0)-i\int_{0}^{t} \dd t' U(t,t')\hat{H}_I(t')U_0(t',0).
	\label{eq:Dyson}
	\end{equation}
		The time evolution operators $U_0(t,t')$, $U_V(t,t')$ and $U(t,t')$ describe propagation for the atomic, Volkov and full Hamiltonians, respectively.
	In the strong-field approximation the full time-evolution operator in the integrand in \eqref{eq:Dyson} can be replaced by the Volkov time-evolution operator, which amounts to neglecting the potential for the ionised electron. The state of our system with this approximation can then be written as
	\begin{equation}
	\ket{\psi(t)}=U_{0}(t,0)\ket{0}-i\int_{0}^{t} \dd t' U_V(t,t')\hat{H}_I(t')U_0(t',0)\ket{0}.
	\label{eq:WF-DysonVolkov}
	\end{equation}
	We write \eqref{eq:WF-DysonVolkov} in the velocity gauge but apply a unitary transformation to the laser interaction Hamiltonian to convert to the length gauge.
	It is possible in \eqref{eq:WF-DysonVolkov} to replace the interaction Hamiltonian with the potential, which ultimately will make the computations simpler \cite{Becker1997}. The Volkov time-evolution operator can be written as
	\begin{equation}
	U_V(t,t')=\int \dd^3 \pb \ket{\psi_V^{\pb}(t)}\bra{\psi_V^{\pb}(t')},
	\label{eq:VolkovOp}
	\end{equation}
	where $\ket{\psi_V^{\pb}(t)}=\ket{\pb+\Ab(t)}e^{-iA_V(\pb,t)}$ and $A_V(\pb,t)=\frac{1}{2}\int \dd t (\pb+\Ab(t))^2$
	The atomic time-evolution operator is given by $U_0(t',0)=\exp(i \ip t')$. Substituting \eqref{eq:VolkovOp} and the atomic time evolution operator into \eqref{eq:WF-DysonVolkov} gives
	\begin{align}
	\ket{\psi(t)}&= \exp(i \ip t)\ket{0}-i\int_{0}^{t} \dd t' \int \dd^3 \pb\:\notag\\
	&\times e^{i A_V(\pb,t)} \ket{\pb+\Ab(t)}\bra{\pb+\Ab(t')}e^{-i A_V(\pb,t')}\hat{V}e^{i \ip t'}\ket{0},
	\label{eq:DysonExplicit}
	\end{align}
	which, with a phase transformation can be written compactly as
	\begin{align}
	\ket{\psi(t)}&=\ket{0}-i\int \dd^3 \pb e^{-i S(\pb,t)} \ket{\tilde{\pb}(t)}M(\pb),
	\intertext{where $\tilde{\pb}(t)=\pb+\Ab(t)$ and} 
		M(\pb)&:=\int \dd t' d(\pb,t') e^{i S(\pb,t')},
		\label{eq:SFA_Mp}
	\end{align}
	with $d(\pb,t')=\braket{\pb+\Ab(t')|\hat{V}|0}$ and $S(\pb,t)=\ip t +\frac{1}{2}\int \dd t (\pb+\Ab(t))^2$.
	
	\begin{figure*}
		\includegraphics[width=1.\linewidth]{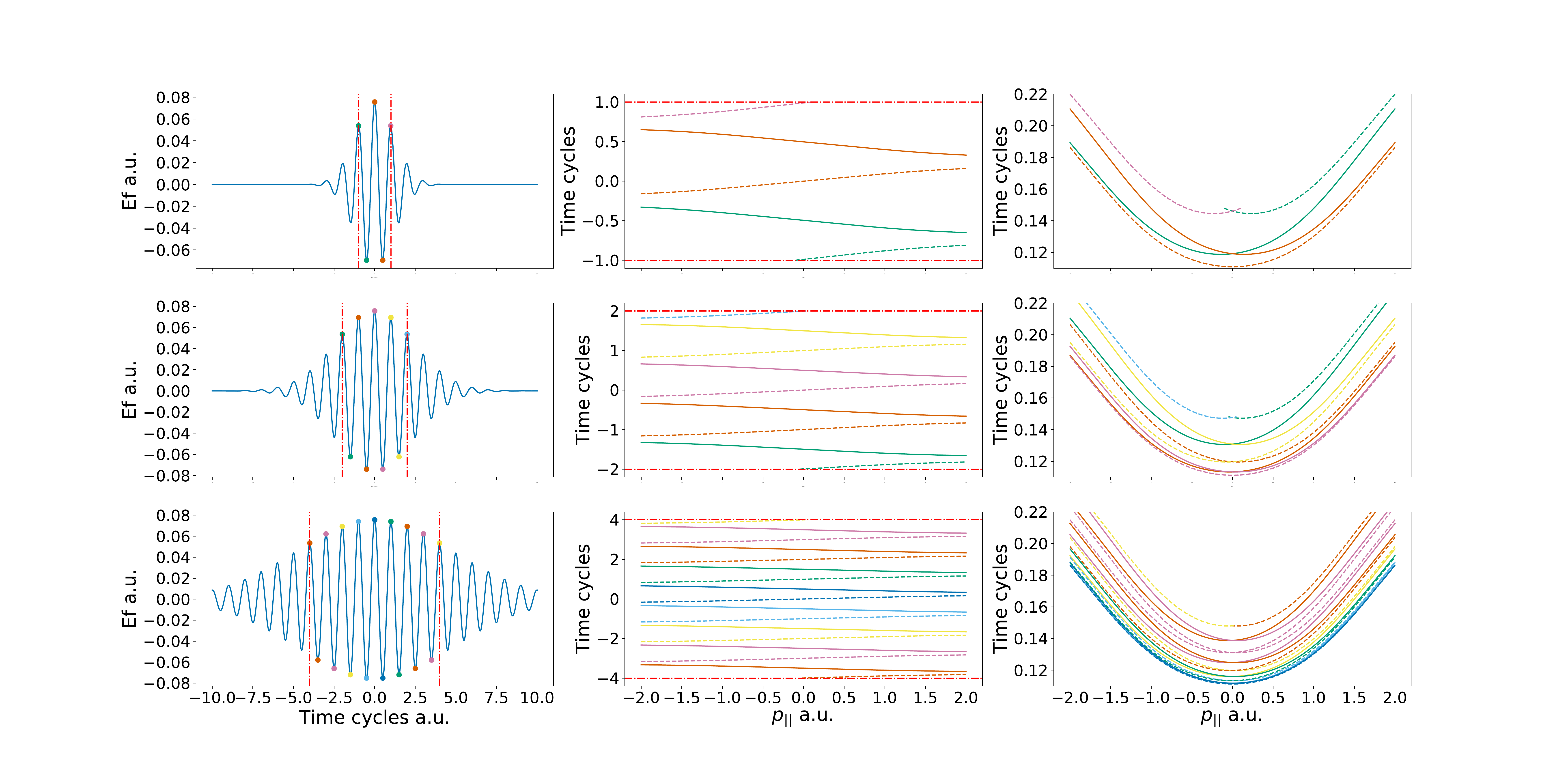}
		\caption{Solutions of the saddle point \eqref{eq:saddleSFA} for 3 Gaussian pulse of 2, 4 and 8 cycles [top (a)--(c), middle (d)--(f) and bottom (g)--(h) rows]. Left column: the real part of the time of ionization on the electric field for a near zero final momentum. Middle column:  real parts of ionization times for varying $p_{||}$ momentum. Right column: The imaginary part of ionization times for varying $p_{||}$ momentum. Dotted lines are drawn vertically to denote the full-width half-maximum of each field. The peak intensity and wavelength of all fields shown is 2$\times$10$^{14}$~W/cm$^2$ [$\up=0.44$~a.u.] and $\lambda = 800$~nm [$\omega=0.057$~a.u.], respectively.
		An ionization potential $\ip=0.5$~a.u.\  is used for the zero range potential.
		The CEPs are all $\phi=\pi/2$, to make the pulses symmetric.
	    The coloured dots mark the real part of solutions of the ionization \eqref{eq:saddleSFA}  on the laser field for a final momentum of $(p_{\parallel},p_{\perp})=(0.002, 0.1)$~a.u. The times of ionization in the middle and right column are plotted using $p_{\perp}=0.1$~a.u., where $p_{\perp}$is the momentum co-ordinate perpendicular to the laser field polarisation. Note the colour used for the dots on the laser field corresponds the colour used for the real and imaginary plots of the solutions in the middle and right columns. 	
		}
		\label{fig:PulseSolutionsExamples}
	\end{figure*}

	In order to calculate the integral over $t'$ we employ the saddle-point approximation and compute stationary action for $t'$ \cite{Faria2002}, which gives	 
	\begin{equation}
	\delta_{t'} S(\pb, t')=0 \implies \left(\pb+\Ab(t')\right)^2=-2 \ip.
	\label{eq:saddleSFA}
	\end{equation}
	The vector potential reads \[\Ab(t)=2\sqrt{\up}F(t)\cos(\omega t + \phi)\hat{e}_{\parallel},\] where $\phi$ is the carrier envelope phase and the pulse envelope $F(t)=e^{-2\ln(2)\frac{t^2}{\tau^2}}$ for a Gaussian pulse with a pulse length $\tau$. For a monochromatic field $F(t)=1$ and $\phi=0$.
	Here, $\hat{e}_{\parallel}$ is the unit vector in the polarization direction of the linear field, often assocaited with the $z$-axis.
	The ponderomotive energy is given by
	\begin{equation}
		\up=\frac{I}{2 c \epsilon_0 \omega^2},
	\end{equation}
	where $c$ and $\epsilon_0$ are the speed of light and vacuum permittivity, given by $c\approx137.036$ and $\epsilon_0=1/(4\pi)$ in atomic units.
	For a monochromatic field \eqref{eq:saddleSFA} is analytically invertible \cite{Maxwell2017} to give the periodic solutions
	\begin{align}
	t'_{en}&=\frac{2\pi (n-e)}{\omega}
	-\frac{(-1)^e}{\omega}\arccos\left(\frac{-p_{\parallel}+(-1)^e i\sqrt{2\ip+p_{\perp}^2}}
	{2\sqrt{\up}}\right),
	\label{eq:TimesMono}
	\end{align}
	where $e=0,1$ denotes pairs of intracycle solutions within the same field cycle and $n=1,2,\dots$ denotes intercycle periodic solutions in each laser cycle.
	In the case of a Gaussian laser pulse \eqref{eq:saddleSFA} must be solved numerically, but the solutions have the same form of intracycle and intercycle solutions.
	
Using the saddle point approximation the transition amplitude can be written as
	\begin{align}
	&M(\pb)= \notag \\
	&\sum^{N_c-1}_{n=0}\sum^{1}_{e=0}\sqrt{\frac{2\pi i}{\partial^2 S(\pb,t'_{en})/\partial {t'}^2} }\braket{\pb+\Ab(t')|\hat{V}|0}e^{i S(\pb,t'_{en})}
	\label{eq:TransitionAmplitudeMom}.
	\end{align}
	For the monochromatic case this can be rewritten as
	\begin{align}
	&|M(\pb)|^2=\notag \\
	&\Omega(\pb)
	\left|
	\sum^{1}_{e=0}
	\sqrt{\frac{2\pi i}{\partial^2 S(\pb,t'_{e0})/\partial {t'}^2} }\braket{\pb+\Ab(t')|\hat{V}|0}e^{i S(\pb,t'_{e0})}
	\right|^2,
	\label{eq:intrainter}
	\end{align}
	where
	\begin{align}
	&\Omega(\pb)=\frac{\cos\left[\frac{2\pi N_c}{\omega} \left(
		\up+\ip+\frac{1}{2}\left|\pb\right|^2\right)\right]-1}
	{\cos\left[\frac{2\pi}{\omega} \left(
		\up+\ip+\frac{1}{2}\left|\pb\right|^2\right)\right]-1}.
	\label{eq:TransitionAmplitudeMom_Pref}
	\end{align}
\eqref{eq:intrainter} states that intracycle and intercycle interference are separable for monochromatic fields \cite{Maxwell2017}. Explicitly, intercycle interference is given by \eqref{eq:TransitionAmplitudeMom_Pref}, which gives a quantization condition for rings of constant energy in the momentum space. 

Note the SFA model presented here is cylindrically symmetric. Thus, it may be full parametrized by the final momentum coordinates $p_{||}$ and $p_{\perp}$, parallel and perpendicular to the laser field polarization, respectively.


\subsection{Features of the Model}
\label{sec:NumericalResults:Features}
\begin{figure}
	\centering
	\includegraphics[width=1.\linewidth]{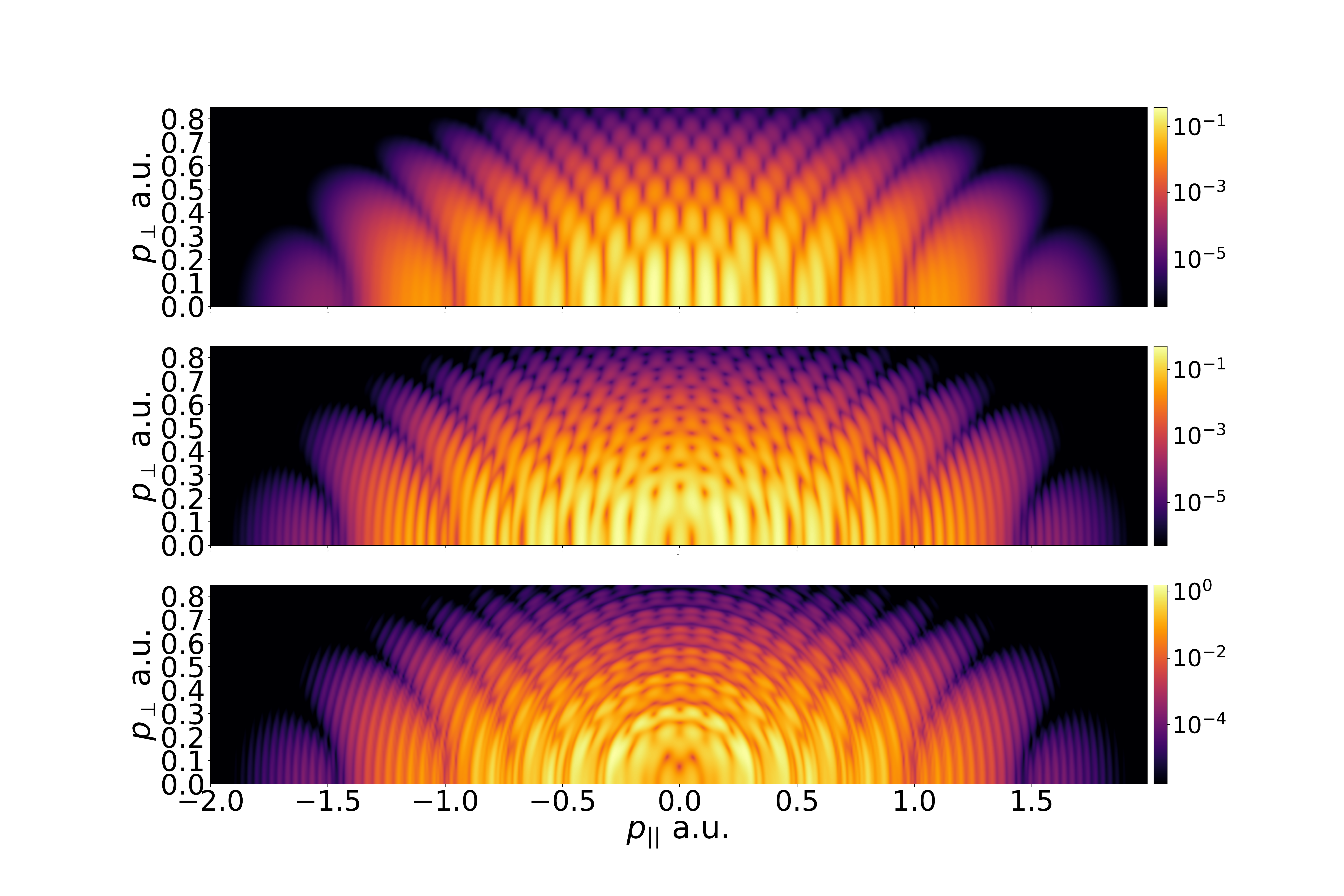}
	\caption{Momentum dependent transition amplitude computed for the same target and field parameters as \figref{fig:PulseSolutionsExamples}, which corresponds to 2, 4 and 8 cycle Gaussian pulses from top to bottom. Further examples of such distributions can be found in \cite{Korneev2012}.
	}
	\label{fig:MomentumDisExamples}
\end{figure}
	
	Before deriving the quantum and classical Fisher information, we first briefly present the possible ionization times that the electron may take via the quantum orbit model for different pulse lengths, and how this affects the momentum dependent transitions amplitude.

	In \figref{fig:PulseSolutionsExamples} and \figref{fig:MomentumDisExamples} the solutions of \eqref{eq:saddleSFA} and transition amplitude show a few well-known physical properties of the system:
	\begin{enumerate}[i.]
		\item The real part of the times of ionization occur at the electric field maxima for $p_{\parallel}=0$ [\figref{fig:PulseSolutionsExamples} (left column)].
		\item Away from $p_{\parallel}=0$ the solutions move slowly towards the field crossings [\figref{fig:PulseSolutionsExamples} middle column].
		\item The imaginary part gives a good indication of the photoelectron yield, with a lower value giving a higher yield [\figref{fig:PulseSolutionsExamples} right column]
		\item The orbits occurring nearest to the electric field maximum have the highest yield.
		\item The overall yield is maximum at $\pb=0$ [\figref{fig:MomentumDisExamples}].
		\item The shorter laser pulses display characteristic finger-like fringes, due to interference between orbits occurring within a single laser cycle. [\figref{fig:MomentumDisExamples} (a)]
		\item The long pulse also demonstrate interference rings between laser cycles, known as above-threshold ionization (ATI) rings. [\figref{fig:MomentumDisExamples} (c)]
	\end{enumerate}
	
	\subsection{The Fisher Information}
	\label{sec:Theory:FisherInformaition}
	 A quantum measurement described by a positive-operator valued measure (POVM) $K_{\mu}\,$:$\,\sum_{\mu} K_{\mu}^{\dag}K_{\mu}={\mathbb{1}}$ allows for the estimation of an unknown parameter $g$, encoded in a quantum state $\varrho_{g}$, with an uncertainty limited by the classical Cram{\'e}r-Rao bound \cite{Helstrom1976,Paris2009}:	
	\begin{equation}
	\Delta g \ge \frac{1}{\sqrt{N I_{F}}} \; ,
	\label{eq:cramer-rao}	
	\end{equation}	
	where $N$ is the number of individual measurements performed and
	$I_F$ is the classical Fisher information, given by \begin{equation}	
	I_F = \sum_{\mu} \frac{\left(\mathcal{P}'(\mu|g)\right)^2}{\mathcal{P}(\mu|g)}
	\label{eq:ClassicalFisher} 	
	\end{equation}
	in terms of the conditional probability $\mathcal{P}(\mu|g)= {\rm Tr}[K^{\dag}_{\mu}K_{\mu}\varrho_g]$ (while a prime $'$ denotes derivation with respect to the parameter $g$). In turn, the classical Fisher information is bounded from above by the quantum Fisher information $Q_F = \sup_{K_{\mu}} I_F$, which may be evaluated as
%
	\begin{equation}
	Q_F(t)=4\left[\braket{\partial_g \psi (t)|\partial_g \psi (t)}
	-\left|\braket{\psi(t)|\partial_g \psi(t)}\right|^2\right],
	\label{eq:QuantumFisher}
	\end{equation}
	where $\partial_g$ is the derivative with respect to the measurement variable, which in our case is the ponderomotive energy $\up$. 
	
	
	\section{Analytical Results}
	\label{sec:AnalyticalResults}
	
	In order to compute both the classical and quantum Fisher information we must calculate the derivative of the wave function with respect to the measurement variable.	
	\begin{align}
		\partial_g\ket{\psi(t)}&=
		-i\int \dd \pb \ket{\tilde{\pb}(t)}e^{-i S(\pb,t)}M_g(\pb,t),
	\intertext{where}
		M_g(\pb,t) &= \int \dd t' f(\pb,t,t') e^{i S(\pb,t')},\\
		f(\pb,t,t') &=\frac{\partial d(\pb,t')}{\partial g}+ i\, d(\pb,t')\left(\frac{\partial S(\pb,t')}{\partial g}
		-\frac{\partial S(\pb,t)}{\partial g}\right).
	\end{align}
	The definition of $d(\pb,t')$ is given in \eqref{eq:SFA_Mp}.
	We are assuming that we are choosing $t$ such that $\Ab(t)=0$ and $\ket{\tilde{\pb}(t)}=\ket{\pb}$. This is achieved when laser the pulse is over or if the vector potential is at a crossing point.	
	The transition amplitude derivative $M_g(\pb,t)$ can be computed using the same saddle points as $M(\pb)$ but just using the alternative prefactor $f(\pb,t,t')$, which saves on computational time.
	
	\subsection{Quantum Fisher Information}
	\label{sec:AnalyticalResults:Quantum}
	Inserting the above definitions for the wavefunction and its derivative in to \eqref{eq:QuantumFisher} gives an expression for the quantum Fisher information in terms of the transitions amplitude and its `derivative', namely
	\begin{align}
		Q_F(t)&=4\left\{
		\int\dd^3 \pb |M_g(\pb,t)|^2 
		-\left|
		\int \dd^3 \pb \conj{M(\pb)}M_g(\pb,t)
		\right|^2
		\right\}.
		\label{eq:QuantumFisherExplicit_Fast}
	\end{align}
There is only one quantum Fisher information as it corresponds to the supremum of the set of classical Fisher informations over all possible POVMs, thus represent the best possible measurement and is basis independent.

	\subsection{Classical Fisher Information}
	\label{sec:AnalyticalResults:Classical}
	In all cases the POVMs corresponding to particular experimental measurements will be on the Hilbert space $L^2(\mathbb{R}^3) \oplus \{\ket{0}\}$ of square integral functions on 3D space, which corresponds to measurements of the momentum (or space) on the ionized portion of the wavefunction plus the zero projector corresponding to the bound portion of the wavefunction. We will use the notation that $K_0=\ket{0}\bra{0}$ throughout.
	Next, we consider various cases of measurement:
	\begin{enumerate}[1.]
		\item An idealized, infinitely precise, momentum measurement, in which the outcomes correspond exactly to the elements in the full momentum space $\pb\in\mathbb{R}^3$. We will denote these measurements as \textit{full} for short.
		\item Coarse momentum measurement, with a tunable precision. The possible outcomes will divide momentum space into a finite number of regions. This provides a more realistic experimental measurement. We will denote these measurements as \textit{coarse}.
		\item A measurement relating to ionization yield, which will only have two outcomes. The whole momentum region will be taken as a single outcome that relates to the electron being ionized, while the bound state will relate to the other outcome and represents the electron remaining bound. These measurements will be  labelled as \textit{yield}.
		\item We will also consider \textit{spectral} measurements, where the energy of the photoelectron is measured but no angular information is collected. This will have a version with infinite precision and a coarse version, that correspond to an infinite and finite set of energy outcomes, respectively. We will refer to the full and coarse spectral measurements as \emph{spec1} and \emph{spec2}, respectively.
	\end{enumerate}
	\subsubsection{Full Measurements}
	For the momentum case we will be measuring with the POVM
	\begin{equation}
		\id 
		:=\,K^{full}_{0}+		
		\int \dd^3 \pb\, K^{full}_{\pb},
		\label{eq:mometumPOVM}
	\end{equation}
	where 
	$K^{full}_{\pb}=\ket{\pb}\bra{\pb}$.
	Recall, the additional bound state projector simply represents the portion of the wavefunction that remains bound, however, this projector is zero when the derivative is taken in the classical Fisher information and thus it does not affect the final result but in other formulations it could play a role. The classical Fisher information takes the form
	\begin{align}
	I^{full}_{F}(t)&=\int \dd^3 \tilde{\pb}(t)\frac{ \left(\partial_g\left|
		-ie^{-i S(\pb,t)}M(\pb)
		\right|^2\right)^2}
	{\left|-ie ^{-i S(\pb,t)}M(\pb) \right|^2}\\
	I^{full}_{F}(t)	&=\int \dd^3 \tilde{\pb}(t)\frac{ \left(\partial_g\left|
		M(\pb)
		\right|^2\right)^2}
	{\left|M(\pb) \right|^2}.
	\intertext{From the above expression it is clear for the momentum basis there is no explicit time-dependence in the classical Fisher information. The expression can further simplify to }
	I^{full}_{F}(t)	&=\int \dd^3 \pb\frac{ 4 \Re\left[M_g(\pb,t)\conj{M(\pb)}\right]^2}
	{\left|M(\pb) \right|^2}.
	\label{eq:ClassicalFisherMomentum}
	\end{align} 
	Note if we remove the time dependence on the integration measure it acts to shift the centre of the integrand but the integration region is infinite. Thus, this does not affect the final result. When we compute this classical Fisher information we will refer to it as the \textbf{full} classical Fisher information.
	
	\subsubsection{Coarse Measurements}
	\label{sec:AnalyticalResults:CoarseMeasurements}
	The POVM given by \eqref{eq:mometumPOVM} represents momentum measurements with infinite precision. In experiment there is some uncertainty associated with the measurements. One way to represent this is with coarse measurement, where we imagine there is a discrete number of momentum bins. They can be described by the following POVM:
	\begin{equation}
		\id		:=\, \sum_{i=0}^{N} K^{coarse}_{i},
		\label{eq:coarse_POVM}
	\end{equation}
	where $K^{coarse}_{i}$ are the projective measurements for each outcome, there are $N$ momentum outcomes/ bins and $K^{coarse}_0=\ket{0}\bra{0}$ and corresponds to portion of the wave function that has not been ionized. The remaining measurement outcomes are described by 
	\begin{equation}
		K^{coarse}_{i} = \int_{\mathcal{R}_i} \dd^3 \pb \ket{\pb}\bra{\pb}.
	\end{equation}
	In this equation $\mathcal{R}_i\subseteq \mathbb{R}^3$ corresponds to a region or bin in momentum space that relates to a single measurement outcome of the detector. Note $\mathcal{R}_i \cap \mathcal{R}_j=\emptyset$ if $i\ne j$ and $\bigcup_{i=1}^{N}\mathcal{R}_i=\mathbb{R}^3$. Inserting this in to \eqref{eq:ClassicalFisher} leads to
	\begin{equation}
		I^{coarse}_{F}(t)=\sum_{i=1}^{N}\frac{4\Re\left[ \int_{\mathcal{R}_i}\!\dd^3\pb  \conj{M(\pb)}M_g(\pb,t)\right]^2
		}{\int_{\mathcal{R}_i}\!\dd^3\pb \left|M(\pb)\right|^2}.
	\end{equation}
	This form of the classical Fisher information we will refer to as the \textbf{coarse} classical Fisher information.
	
	\subsubsection{Yield Measurements}
	\label{sec:AnalyticalResults:RateMeasurements}
	
	The limiting case of coarse measurement is when there is only a single region for the momentum $\mathcal{R}_1=\mathbb{R}^3$. In this case there are only two measurement outcomes, labelled $0$ and $1$, which correspond to if the electron is bound or ionized, respectively, given by the following POVM
	\begin{align}
	\id\,&:=\, K^{yield}_{0} + K^{yield}_{1},
		\intertext{where}
		K^{yield}_{1}&=\int_{\mathbb{R}^3} \dd^3 \pb \ket{\pb}\bra{\pb}.
	\end{align}
	This leads to 
	\begin{equation}
		I^{yield}_F(t)=\frac{4\Re\left[ \int\!\dd^3\pb  \conj{M(\pb)}M_g(\pb,t)\right]^2
		}{\int\!\dd^3\pb \left|M(\pb)\right|^2}
	\label{eq:FisherRate}
	\end{equation}
	for the classical Fisher information. When we use this classical Fisher information we will refer to it as the \textbf{yield} classical Fisher information.
	
	\subsubsection{Spectral Measurements}
	\label{sec:AnalyticalResults:SpectralMeasurements}
	
	It is common in experiment to measure only the energy of the photoelectrons discarding the angular information.
	The spectral measurement using an infinite number of energy bins is given by
	\begin{equation}
		I^{spec1}_F(t)=\int \dd E \frac{\mathcal{P}'(E|g)^2}{\mathcal{P}(E|g)},
	\end{equation}
	while the formula for a coarse measurement with a finite number of energy bins is given by
	\begin{equation}
		I^{spec2}_F(t)=\sum_{i=1}^{N} 
		\frac{\left[\int_{E_{i-1}}^{E_{i}} \dd E\; \mathcal{P}'(E|g)\right]^2}
		{\int_{E_{i-1}}^{E_{i}} \dd E\;\mathcal{P}(E|g)}.
	\end{equation}
	We will refer to these as the full spectra and coarse spectral classical Fisher informations, respectively.
	In both of these formulae $\mathcal{P}(E|g)$ is given by
	\begin{equation}
	\mathcal{P}(E|g) = \int \dd\Omega |M(E,\theta,\phi)|^2,
	\end{equation}
	where $M(\pb)$ has been re-parametrized in terms of the energy and angular dependences and is integrated over the angles.
	
	\subsection{Analytical trends}
	\label{sec:AnalyticalTrends}
	
	\subsubsection{Ponderomotive phase QF}
	\label{sec:AnalyticalTrends:PonderomotivePhase}
	
	In this section we argue that our measurement variable, the laser intensity, is primarily encoded via a phase difference (the ponderomotive phase) between the continuum and bound part of the wave function. It is important to note that if the photoelectron was prepared into a single continuum state with a phase that depended on the laser field intensity this would not be a physical observable and the quantum Fisher information would be zero. However, the ionization process leads to a superposition/ distribution across many continuum states and the bound state, each with their own momentum dependent phases and normalizations. This superposition of final states with different normalizations and phases (relative to the bound state) encoding the laser intensity leads to a large quantum Fisher information, and we show here this phase is quadratically dependent on the time the photoelectron is exposed to the laser field.
	
	Using \eqref{eq:QuantumFisherExplicit_Fast} we can extract an approximate expression for the time dependence of the Fisher information. The transition amplitude $M(\pb)$ is independent of the final time $t$ after the lase pulse is over due to the momentum conservation. It is only $M_g(\pb,t)$ that is dependent on final time via the prefactor term $f(\pb,t,t')$. We can compute the derivative 	
	\begin{align}
			\frac{\partial S(\pb,t)}{\partial {\up}} &= -\frac{1}{2{\up}}
		\left[\pb\cdot\int\Ab(t)\dd t+ \int \Ab(t)^2\dd t\right]
	\end{align}
	of the action in this prefactor, which is the part dependent on the final time.
	For large enough time $t$ the $\Ab(t)^2$ integral in the prefactor will strongly dominate. Thus, the large time behaviour of the Fisher information can be derived as
	\begin{align}
		f(\pb,t,t')&\rightarrow i \frac{d(\pb,t')}{2 \up}\int \Ab(t)^2 \dd t\\
		M_g(\pb,t)&\rightarrow i \int \Ab(t)^2 \frac{M(\pb)}{2\up}\\
		Q_F(t) &\rightarrow \alpha \left(\frac{1}{\up}\int \Ab(t)^2 \dd t\right)^2,
		\intertext{where}
		\alpha &= 
		\int\dd^3 \pb \left|M(\pb)\right|^2 \left(1-\int\dd^3 \pb\left|M(\pb)\right|^2\right).
	\end{align}	
	For a monochromatic field
	\begin{align}
			Q_F(t)&\approx \alpha\left(2 t + \frac{\sin(2 \omega t)}{\omega \up}\right)^2,
			\label{eq:QuantumFisherLimit}
	\end{align}
	which means there will be quadratic scaling in the uncertainty. In the case of a Gaussian pulse
	\begin{align}
	Q_F(t)&\approx \alpha\left(\sqrt{\frac{\pi}{4 \ln(2)}} \tau\right)^2 \Bigg(\erf\left(\frac{t \sqrt{4 \ln(2)}}{\tau}\right) \notag\\
	&+e^{-\frac{\tau^2\omega^2}{4\ln(2)}}	f_{osc}(t)
	\Bigg)^2,
	\intertext{where}
	f_{osc}(t)&=e^{i 2\phi}\erf\left(\frac{4\ln(4) t-i\tau^2\omega}{2\tau \sqrt{\ln(2)}}\right)
	+\mathrm{c.c.}
	\end{align}
	and $\tau$ is the pulse length. Here $f_{osc}(t)$ is oscillatory and heavily damped by the exponential prefactor. The pulse length puts a limit on the quantum Fisher information. For $t\rightarrow\infty$ this expression becomes
	\begin{align}
	Q_F(t)&\propto \frac{\tau^2 \pi \alpha}{8 \ln(2)}
	\left( 1+\cos(\phi)e^{\frac{-\tau^2\omega^2}{2 \ln(2)}} \right)^2.
	\end{align}
	
	It is clear that the quantum Fisher information is quadratically dependent on the laser pulse length rather than final time. For continuous wave/ monochromatic fields the pulse length is in theory infinite so it is the final time/ time until measurement $t$ that determines how long the photoelectron interact with the field for. This quadratic dependence on the integral of the vector potential enables highly sensitive measurements of the laser intensity. We will call the $A^2$ phase difference between the bound and continuum parts of the wave function the ponderomotive phase, as it derives from the electron dynamics in the continuum for a free electron. 	
	In order to measure this phase directly one would need to find a basis in which the bound and final states can interfere. In practical terms, this can be achieved by considering the two pathways, where the electron ionization occurs at two separate times. While one path is ionized and the other is bound a phase difference, via the ponderomotive phase, will be acquired.
	This naturally occurs in strong field phenomena where at various times a portion of the electron wavefunction may ionize. The ionized portion will pick up the ponderomotive phase, while the bound portion will not. Thus, the interference of two electron pathways released at different times, such as in ATI rings, can reveal the ponderomotive phase.
	
	\subsubsection{ATI rings and classical Fisher information}
	In contrast to the quantum Fisher information the classical Fisher information is dependent on the measurement basis.
	Furthermore, if performing momentum measurements the wave function's superposition over the bound and continuum states will not reveal the ponderomotive phase. It requires the interference of more than one electron path with different phases that end in the same continuum state. In this section we will examine how interference affects the classical Fisher information.
	The interference of an electron ionized at two different times of equivalent laser intensity $|\mathrm{E}(t_1)|=|\mathrm{E}(t_2)|$ gives the following probability
	\begin{align}
		|M(\pb)|^2&=|M_1(\pb)+M_2(\pb)|^2\notag\\
		&=|M_1(\pb)|^2|1+e^{i\Delta S(\pb,t_2,t_1)}|^2.
	\end{align}
	In order to compute the classical Fisher information one needs to take the derivative
	\begin{align}
		\partial_{\up}(|M(\pb)|^2)&=|1+e^{i\Delta S(\pb,t_2,t_1)}|^2\partial_{\up}(|M_1(\pb)|^2)
		\notag\\&
		-2 \sin(\Delta S(\pb,t_2,t_1))\frac{\partial \Delta S(\pb,t_2,t_1)}{\partial \up} |M_1(\pb)|^2,
	\end{align}
	with regard to the measurement variable $\up$,
	where
	\begin{align}
	\frac{\partial \Delta S(\pb,t_2,t_1)}{\partial {\up}} &= -\frac{1}{2{\up}}
	\left[\pb\cdot\int_{t_1}^{t_2}\!\Ab(t)\dd t+ \int_{t_1}^{t_2} \!\Ab(t)^2\dd t\right].
	\end{align}	
	The same dependence on the ponderomotive phase is apparent. The phase grows with a larger difference between ionization times, which means longer pulses will allow a larger difference in times thus increasing the momentum measurement classical Fisher information via interference.	
	For a monochromatic field we can explicitly examine the dependence on the pulse length. The transition amplitude can be written as
	\begin{equation}
		|M_N(\pb)|^2=\Omega_N(\pb)|M_0(\pb)|^2,
	\end{equation}
	where $M_N(\pb)$ is the transition amplitude and $N$ denotes the number of laser cycles we consider ionization events for. $M_0(\pb)$ is the ionization events across a single cycle  unit cell and $\Omega_N(\pb)$ is given  by \eqref{eq:TransitionAmplitudeMom_Pref}. The derivative of the previous expression used in the classical Fisher information is
	\begin{align}
		\partial_{\up}(|M_N(\pb)|^2)&=\partial_{\up}(\Omega_N(\pb))|M_0(\pb)|^2\notag\\
		&\quad +\Omega_N(\pb)\partial_{\up}(|M_0(\pb)|^2).
	\end{align}
	We can compute the derivative of $\Omega_N(\pb)$
	\begin{align}
		\partial_{\up}(\Omega_N(\pb))&=\chi_N(\pb)\Omega_N(\pb),
	\end{align}
	with
	\begin{align}
		\chi_N(\pb)&=
		\frac{2\pi}{\omega}\left(
		\cot\left(\frac{\pi}{\omega}(\ip+\up+\frac{1}{2}\pb^2)\right)
		\right.\notag\\
		&\left.-N
		\cot\left(\frac{N\pi}{\omega}(\ip+\up+\frac{1}{2}\pb^2)\right)
		\right).
	\end{align}
	The factor of $N$ in the second term will lead to an $N^2$ dependence in the classical Fisher information, which mirrors the quadratic dependence on the pulse length in the quantum Fisher information.
	
	\section{Numerical Results}
	\label{sec:NumericalResults}

	\subsection{General Trends}
	\label{sec:NumericalResults:GeneralTrends}
	When using momentum \textit{in situ} measurements to determine the laser intensity it is expected that interference and in particular ATI-rings will take a prominent role as shown analytically in \secref{sec:AnalyticalTrends}. Using an orbit based method we are able to turn interference on/ off at will and thus we have the ability to examine the effect this has on the quantum and classical Fisher information.

	\subsubsection{Yield vs Interference}	
\begin{figure}
		\centering
		\includegraphics[width=0.995\linewidth]{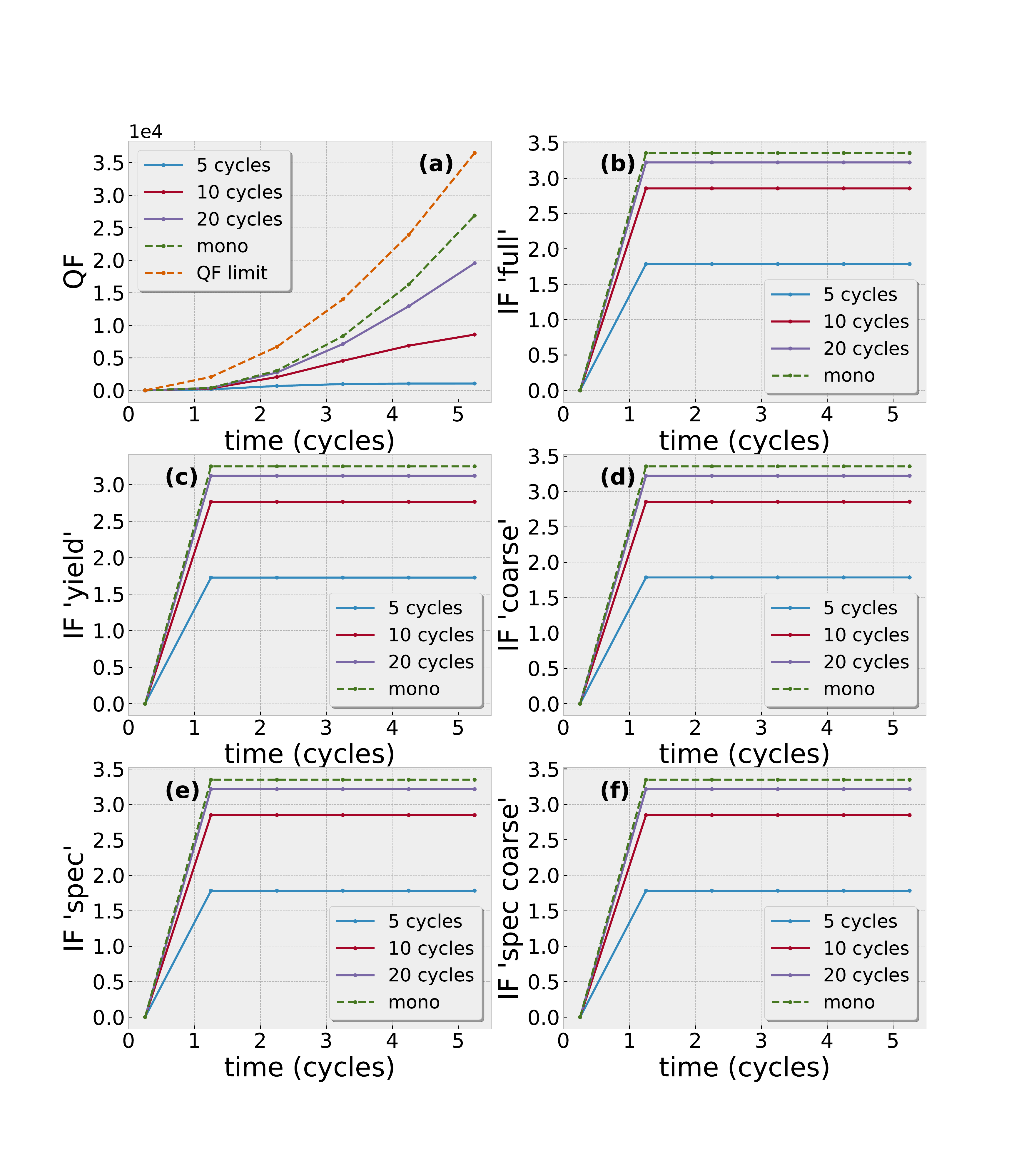}
		\caption{The quantum (a), classical full momentum basis (b), yield basis (c), coarse basis (d), full spectral (e) and coarse spectral (f) Fisher information for a single ionisation channel. The computation has been done for three Gaussian pulses each with full-width at half maximum of $5$, $10$ and $20$ laser cycles and a monochromatic laser for reference. 
		The peak intensity and wavelength of all fields shown is 2$\times$10$^{14}$~W/cm$^2$ [$\up=0.44$~a.u.] and $800$~nm [$\omega=0.057$~a.u.], respectively.
		An ionization potential $\ip=0.5$~a.u.\  is used for the zero range potential.
		The CEPs are all $\phi=\pi/2$, to make the pulses symmetric.
		The coarse momentum and spectral measurements use a resolution of $0.1$~a.u.\  and $0.05$~a.u.\ respectively. In the case of the coarse momentum measurement this means the regions/ bins $\mathcal{R}_i$ from \eqref{eq:coarse_POVM} are squares with sides of $0.1$~a.u.}
		\label{fig:Rate_Trends}
	\end{figure}

	Earlier measurements simply used the ionization yield to determine the laser intensity. In order to replicate these results we start by computing the quantum and classical Fisher information using only a single ionisation channel/ orbit so that all interference effects are removed, see \figref{fig:Rate_Trends}. This is a little artificial as we consider three different pulse lengths and a monochromatic field but restrict them all to a single ionisation event. This makes both the quantum and classical Fisher information smaller but allows us to probe trends without interference effects.
	 
	The quantum Fisher information is not affected by the lack of interference and as expected from \secref{sec:AnalyticalTrends:PonderomotivePhase}. The quantum Fisher information quadratically increases with time for the monochromatic case (see \figref{fig:Rate_Trends} (a)) and for the Gaussian pulse it follows an error function, reaching a maximum at the end of the laser envelope. This is consistent with the freed electron acquiring laser intensity information for as long as it is exposed to the laser, via a ponderomotive phase. The quantum Fisher information effectively gives the maximum classical Fisher information for any measurement basis. Thus, measuring in some non-momentum  bases the superposition of the bound state with the continuum states can be exploited via interference to reveal their phases.	
	The limit given by \eqref{eq:QuantumFisherLimit} for the quantum Fisher information is plotted on \figref{fig:Rate_Trends} and closely follows the computed value.
	
	In contrast, the classical Fisher information increases from zero once the electron is ionized but remains constant beyond this point. Momentum measurements are blind to the ponderomotive phase if interference is neglected and give information related only to the ionization yield. In addition, the canonical momentum is conserved making the transition amplitude and thus the classical Fisher information constant after ionization.	
	\begin{figure}
		\centering
		\includegraphics[width=\linewidth]{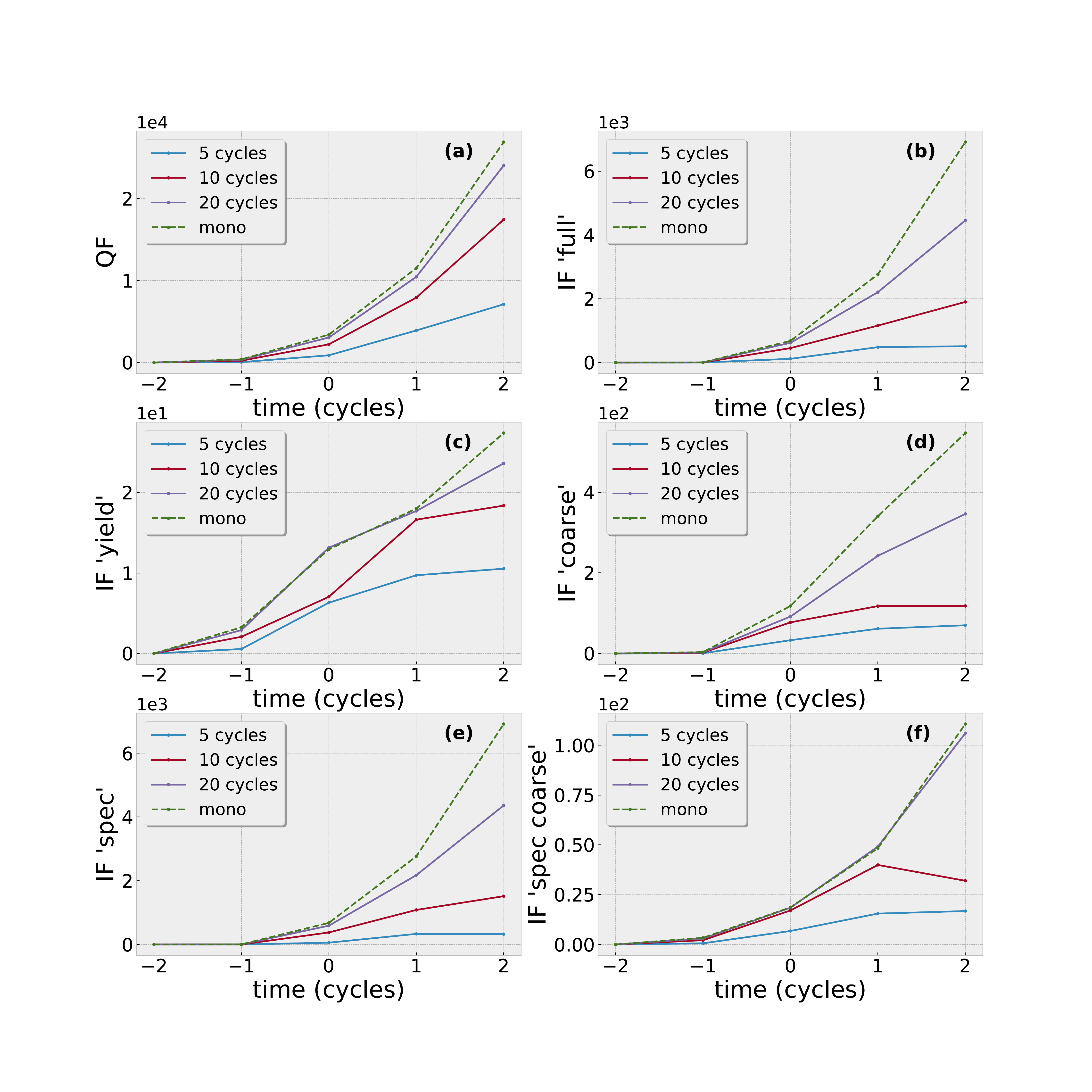}
		\caption{The quantum and classical Fisher informations as in \figref{fig:Rate_Trends} except computed with 5 channels of ionization included to asses the effect of interference.
			The same field and target parameters are used as in  \figref{fig:Rate_Trends}.}
		\label{fig:ATI_Trends}
	\end{figure} 
	
	There are nearly 4 orders of magnitude difference between the classical Fisher informations and quantum Fisher information. With the interference switched off all the classical Fisher informations are equal because only the overall yield contributes to the Fisher information. So the coarse, full and spectral based measurements are all reduced to the value of the yield measurement. The yield measurement will be always a lower bound for the other measurements and thus can be used to quantify how much interference enhances other measurements. 
	The fact that the quantum is so much larger than classical Fisher information suggests the momentum basis is not a good choice for determining laser intensity for strong field ionization if there is little or no interference. This could have implications for very short laser pulses with full-width at half maximum of around 1 cycle. It would suggest there exists a better measurement basis built from a superposition of momentum states. In fact, it is possible that the position basis for the photoelectron would outperform the momentum basis in determining the laser intensity uncertainty for very short pulses.

	In contrast, \figref{fig:ATI_Trends} includes 5 interference channels. These are all relevant channels for the 5 cycle pulse, while for the longer pulses we consider the first 5 relevant channels by order of occurrence in the pulse. This leads to large differences between the classical Fisher informations. As the time increases along the pulse, additional ionization events occur, each acting to increase the value of the various classical Fisher informations.
	The full classical Fisher information drastically increases quadratically with each ionization event to around 3 orders of magnitude larger than the value for a single channel. In contrast, the yield classical Fisher information does not increase very much, about 8 times the value for a single channel. However, 4 extra channels lead to more ionization yield, hence some increase in yield classical Fisher information is expected. The coarse classical Fisher information follows a behaviour roughly in-between the other two cases, increasing just over 2 orders of magnitude times the value for a single channel.
	
	In addition, we have included the classical Fisher information, which results from making `spectral' energy measurement upon the photoelectron. For longer 20-cycle and monochromatic cases this is almost identical to the full momentum measurement. Here we consider only intercycle `ATI' interference. Thus, for long pulses the momentum dependent transition amplitude is almost spherically symmetric hence the similarity between the two. However, for shorter pulses the distributions are less symmetric and the spectral measurement performs less well. For coarse measurement with an energy resolution of $0.05$~a.u.\  the spectral measurements are consistently 5$\times$ worse than the coarse momentum measurements with a resolution of $0.1$~a.u. Note this level of momentum resolution is routinely achievable in experiments \cite{Wolter2015}.
	Aside from the spectral measurements neglecting angular information, it is likely this difference is due to the constant momentum resolution vs constant energy resolution, for which there is the following relationship $\Delta E = 1/2 \Delta p^2+p\Delta p$. Thus, for the region $p\in[0,1]$~a.u., where the majority of signal is located, the fixed $0.1$ a.u.\ momentum resolution varies from $0.005 - 0.105$ a.u.\ resolution in energy. This gives a lower average resolution of this region but a higher resolution in the region where the distribution is maximum, and thus leads to a lower uncertainty than that of the spectral measurement.	 
	The resolution of detection setups used in strong field physics does not exhibit constant resolution as shown here but will vary quantifiably, often dependent on tuning the guiding electric and magnetic fields \cite{Wolter2015}.

	\begin{table}
		\begin{tabular}{l|rrrr}
			\hline
			{} &          5 &         10 &         20 &       Mono \\
			\hline
			full        &  24.328822 &  54.805266 &  77.883953 &  89.462169 \\
			yield        &  24.742480 &  55.700743 &  79.153587 &  90.920518 \\
			coarse      &  24.336157 &  54.821454 &  77.906904 &  89.488510 \\
			spec        &  24.347465 &  54.860459 &  77.967933 &  89.560844 \\
			spec coarse &  24.352390 &  54.870148 &  77.981217 &  89.575920 \\
			\hline
		\end{tabular}
		
		\caption{Increase in uncertainty of the classical vs quantum Fisher information for the full momentum, yield, coarse momentum, spectral and coarse spectral measurements, neglecting interference. Computed using $\sqrt{Q_F(t_f)/I_F(t_f)}$, where $t_f$ is the time at the end of the pulse. This relates to using only a single channel for ionization. We choose the ionization event that occurs at the peak intensity of the laser. All parameters are the same as \figref{fig:Rate_Trends}.}
		\label{Tab:Uncertainty_Rate}
	\end{table}

	\begin{table}
		\begin{tabular}{l|rrrr}
			\hline
			{} &          5 &         10 &         20 &       Mono \\
			\hline
			full        &   3.727896 &   3.029688 &   2.322227 &   1.972497 \\
			yield        &  25.965643 &  30.805221 &  31.886580 &  31.337388 \\
			coarse      &  10.072111 &  12.156392 &   8.328323 &   7.011814 \\
			spec        &   4.685787 &   3.392254 &   2.346648 &   1.971997 \\
			spec coarse &  20.580409 &  23.347627 &  15.053142 &  15.591103 \\
			\hline
		\end{tabular}
		
		\caption{Same as \tabref{Tab:Uncertainty_Rate} except 5 intercycle channels of ionization are included. These are the first 5 ionization events that occur after the half maximum of each pulse, $-\tau/2$. Any relevant events after this are not considered in these results. All parameters are the same as \figref{fig:ATI_Trends}.}
		\label{Tab:Uncertainty_ATI}
	\end{table}
\begin{table}
	\begin{tabular}{l|rrrr}
		\hline
		{} &          5 &         10 &         20 &       Mono \\
		\hline
		full        &   2.627620 &   2.231962 &   1.993563 &   1.923594 \\
		yield        &  56.859871 &  32.893725 &  22.353922 &  31.975848 \\
		coarse      &   7.583439 &   9.502632 &   7.371176 &   6.649124 \\
		spec        &   4.431234 &   3.810768 &   2.234443 &   1.951190 \\
		spec coarse &  20.225071 &  23.308732 &  16.617459 &  14.007984 \\
		\hline
	\end{tabular}
		\caption{Same as \tabref{Tab:Uncertainty_ATI} except both pairs of ionization times are included for the all the 5 channels of ionization. All remaining parameters are the same as \figref{fig:Rate_Trends}.}
\label{Tab:Uncertainty_ATI2}
\end{table}

	In \tabref{Tab:Uncertainty_Rate} the square root of the ratio between the quantum and the classical Fisher informations for a single ionization channel are displayed. This can be viewed as denoting how close the uncertainty derived from a specific classical Fisher information is compared to the best possible measurement, where a value of 1 would mean they are equal. Note, for \tabref{Tab:Uncertainty_Rate}, the longer the pulse is the larger the ratio, this is because the quantum Fisher information will increase more rapidly for a longer pulse, while the classical Fisher information does not significantly increase when interference is neglected. The situation drastically changes when interference is included in \tabref{Tab:Uncertainty_ATI}, the full classical Fisher information is much closer to the quantum case. With the uncertainty only $3.7$ times minimum possible value for the 5 cycle pulse, reducing to $2.0$ times this for the monochromatic case. The yield classical Fisher information stays at around $30$ times the best minimum possible uncertainty, thus is $7.5 - 15$ less precise than the full classical Fisher information. The coarse classical Fisher information (with $0.1$ a.u. resolution) is in-between at around 9 times the minimum uncertainty for a 5 cycle pulse. The complete spectral measurement is close to full momentum measurement, again due to the near spherical symmetry of the transition amplitude, which increases for longer pulses. The coarse spectral measurement for a resolution of $0.05$~a.u. consistently leads to an uncertainty, twice that of the coarse momentum measurement. This is consistent with the factor of 5 observed between the classical Fisher informations.
	
	In \tabref{Tab:Uncertainty_ATI2} both the `long' and `short' ionization times are included for all 5 channels of ionization. Hence, both intercycle and intracyle interference is included. This leads to further improvement in the full and coarse momentum measurement, as they get closer to the optimal quantum measurement, particularly for the 5 cycle pulse. The spectral measurement, however,  remains at a similar value. This is due to the fact that there will be more angular dependence in the momentum distribution, which spectral measurement will not have access to. Thus, it is likely that more detailed intracyle interferences such as holographic interference patterns, which are not considered here, would lead to even larger differences between the full and spectral measurements. 
	
	This suggests that for a short 5 cycle pulse changing the measurement from a $0.05$~a.u resolution spectral measurement to an arbitrarily high resolution full momentum measurement could reduce the laser uncertainty by $7.7$ times. Considering the state  of the art is around $1\%$ \cite{Pullen2013,Wang2019} uncertainty achieved for a spectral measurement with similar resolution, this could mean uncertainty lower than $0.13\%$. However, these experiments used different pulse length of 2-3 cycles in Ref.~\cite{Pullen2013} and 15 cycles in Ref.~\cite{Wang2019}. In the former a better estimate can be found by using only 3 cycles (see \secref{sec:NumericalResults:Predicitons} \tabref{tab:PullenComparison}). The latter experiment has a pulse long enough and strong enough such that depletion of the bound state should be accounted for. This has not been considered in the current framework. Thus, we must keep the probability of remaining in the bound portion of the wavefunction near one and uncertainty derived from long or high intensity pulses should be treated with caution. This is discussed in more detail in \secref{sec:PulseLengthIntensity}.
	
	To further reduce the uncertainty one could change the experimental setup entirely to measure the ponderomotive phase (in line with the quantum Fisher information), which reduces the uncertainty by an extra factor of $2.6$. Of course, simply doing more measurements will also reduce the uncertainty. However, a reduction by a factor of $10$ requires $100$ times more measurements. This quickly becomes difficult as various experimental parameters can not be controlled/ kept stable enough for extended periods of time, thus there is limit on the acquisition time, see \secref{sec:PulseLengthIntensity} for more details.
	
	\subsubsection{Coarseness}
\begin{figure}
		\centering
		\includegraphics[width=1.05\linewidth]{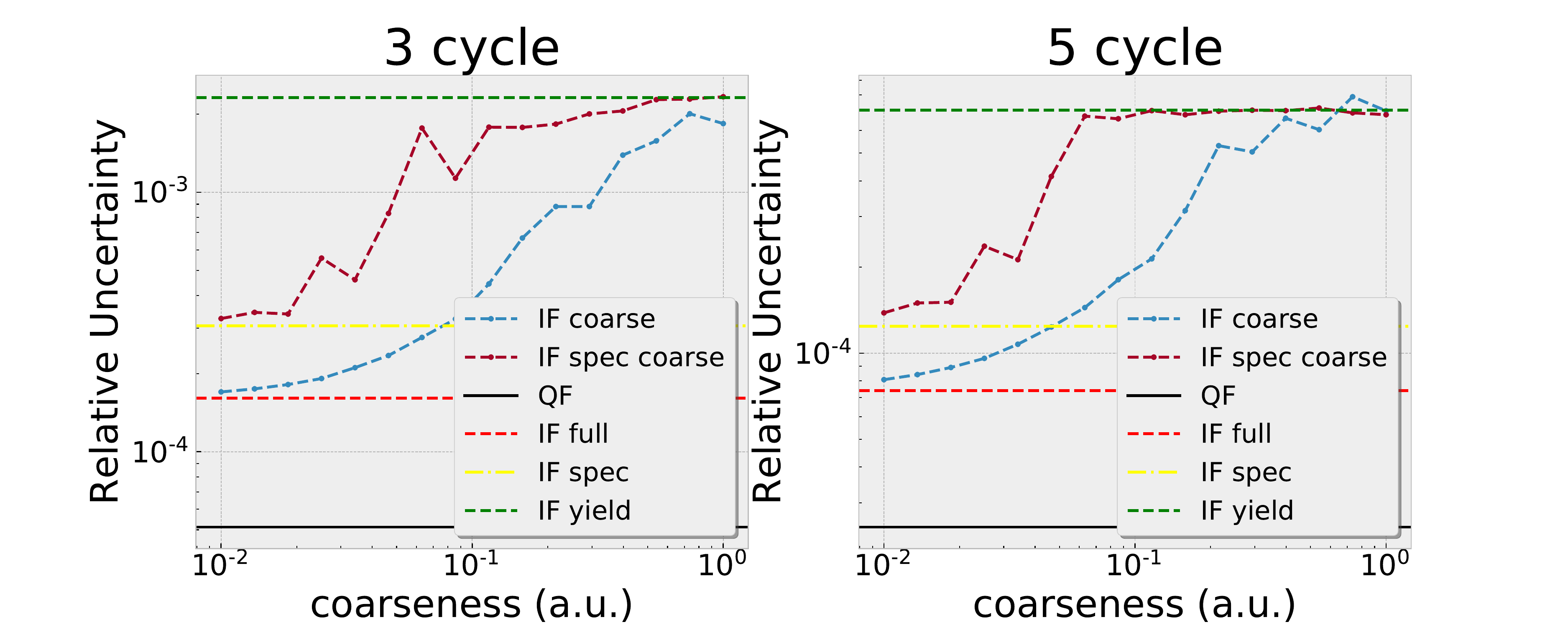}
		\caption{The uncertainty in laser intensity for coarse measurements with different resolutions. On the left is a 3 cycle pulse, on the right is a 5 cycle pulse. The laser intensity is $2\times 10^{14}$~W/cm$^2$.}
		\label{fig:coarsenesscomparison}
	\end{figure}

	In \secref{sec:AnalyticalResults} we presented the formulation for the full, coarse and yield measurements. The full and yield measurements are opposite limits of the coarse case, where full corresponds to infinite precision measurements ($\Delta p \to 0$ or $\Delta E \to 0$) and yield corresponds to no precision ($\Delta p \to \infty$ or $\Delta E \to \infty$).	
	In \figref{fig:coarsenesscomparison} we examine the effect of reducing the coarseness from $1.$~a.u. to $0.01$ a.u. (in momentum and energy resolution) on the relative uncertainty $\Delta \up/\up$, where $\Delta \up$  is given by the Cram\'ers-Rao bound (\eqref{eq:cramer-rao}). 	
	As expected from the above arguments, the relative uncertainty asymptotically tends to the value derived from the full classical Fisher information. 
		
	It is of note that, for the parameters chosen here, increasing the precision only reduces the laser intensity uncertainty [\figref{fig:coarsenesscomparison} (a)] up to a point. Increasing precision from $1-0.1$~a.u. leads to an order of magnitude decrease in uncertainty but changing the precision from $0.1 - 0.01$~a.u. leads only to a factor of two decrease in uncertainty. By $0.01$ precision the value of IF coarse is nearly that of IF full so increasing the precision further will not appreciably reduce the laser intensity uncertainty. Thus, there are diminishing returns to increasing the momentum precision. It does mean that the uncertainty presented here, derived from full momentum measurements, should be realistically achievable in experiment as $0.01$~a.u. resolution is not much beyond that achieved in existing setups \cite{Wolter2015}.

	\subsubsection{Pulse Length and Laser Intensity Trends}	
	\label{sec:PulseLengthIntensity}
	\begin{figure}
		\centering
		\includegraphics[width=1.0\linewidth]{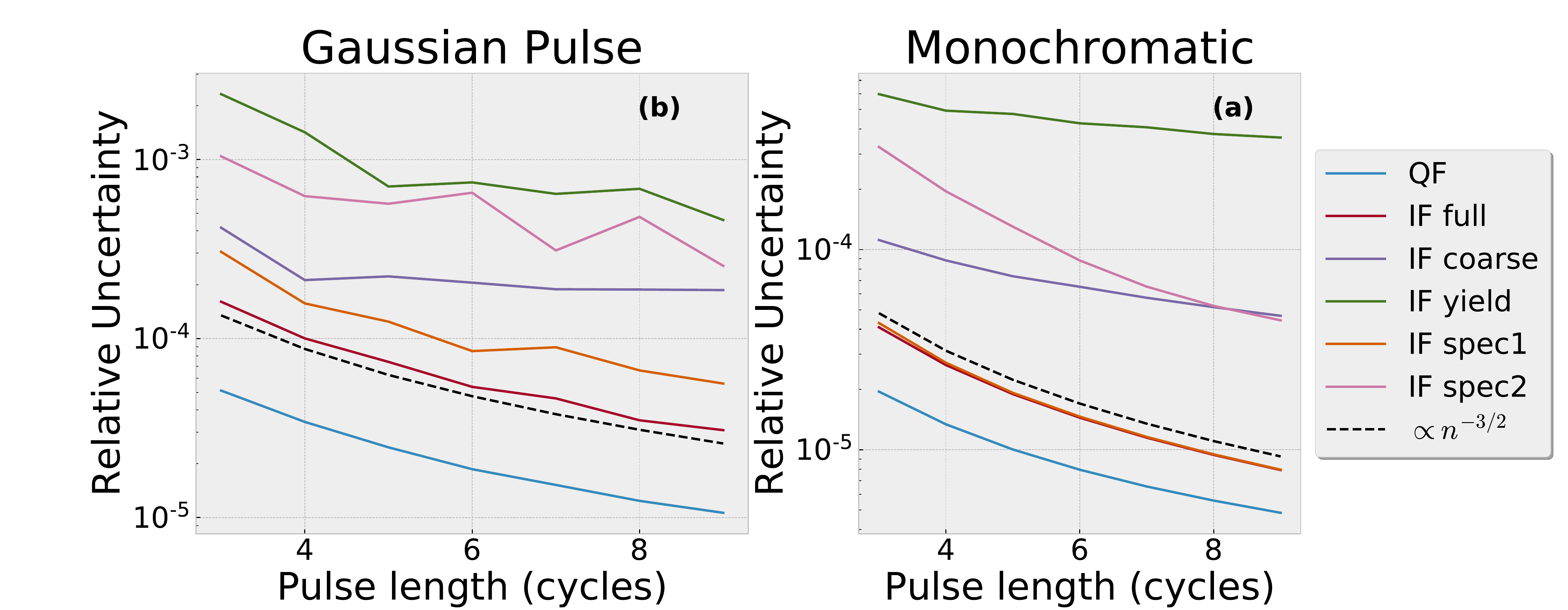}
		\caption{The quantum and classical Fisher information with an increasing number of laser cycles. In the monochromatic case, the number of cycles corresponds to the number of ionization events included. The laser intensity is $2\times 10^{14}$~W/cm$^2$. The remaining field and target parameters are the same as \figref{fig:Rate_Trends}.}
		\label{fig:atitrendsmonovspulse}
	\end{figure}
	\begin{figure}
		\centering
		\includegraphics[width=1.0\linewidth]{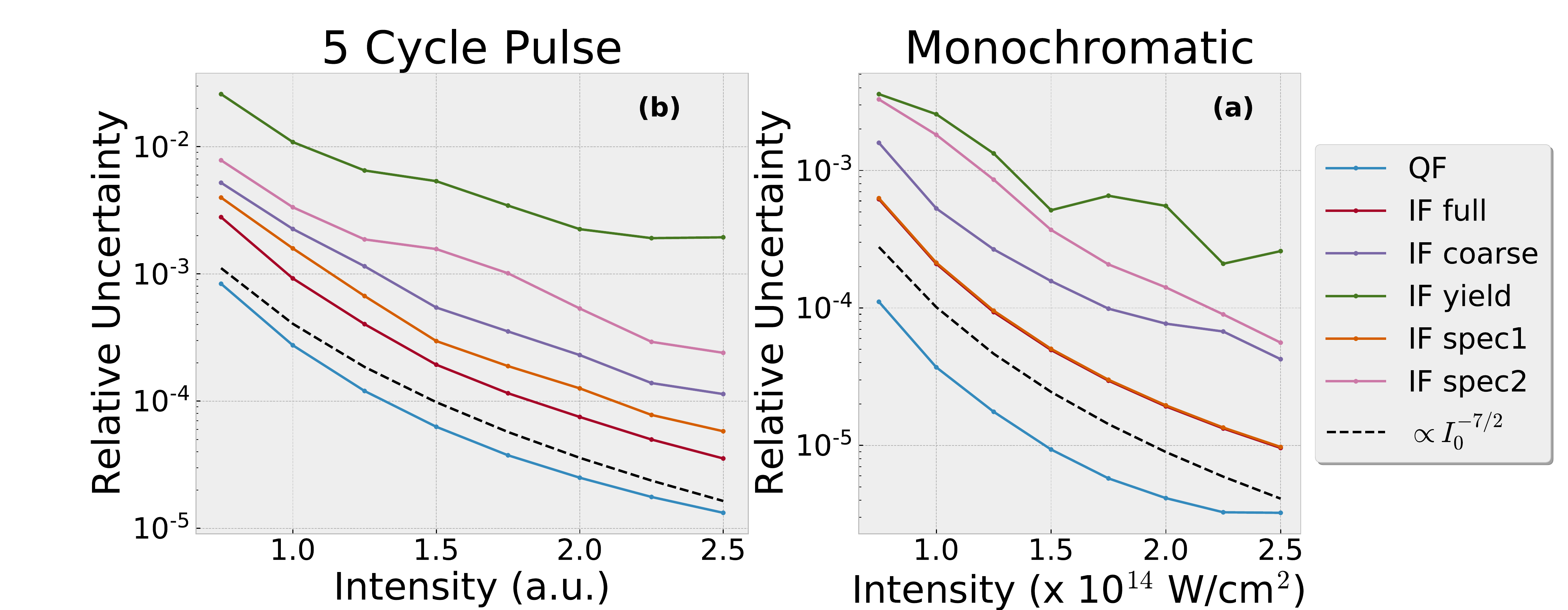}
		\caption{Uncertainty with increasing laser intensity derived from the quantum and classical Fisher informations for a 5 cycle and monochromatic laser field. The remaining field and target parameters are the same as \figref{fig:Rate_Trends}.}
		\label{fig:fullupincrease0}
	\end{figure}	
	
	In \figref{fig:atitrendsmonovspulse} the relative laser intensity uncertainty is plotted vs an increasing pulse length. All ionization channel intracycle and intercycle have been included for each pulse. Note that the relative uncertainty derived from full classical Fisher and quantum Fisher information both follow $n^{-3/2}$ scaling over the number of cycles $n$. This corresponds to a cubic scaling in the respective Fisher informations.
	This is due to a combination of quadratic scaling with the pulse length due to interference combined with additional linear scaling as an increasing the pulse length linearly increases the signal/ portion of the electron that is ionized.
	It is clear from the previous analytic and numerical results that the quantum Fisher information quadratically increases with the time that the laser is on, while the full classical Fisher information follows the same trend due to interferences.


	\begin{figure}
		\centering
		\includegraphics[width=0.9\linewidth]{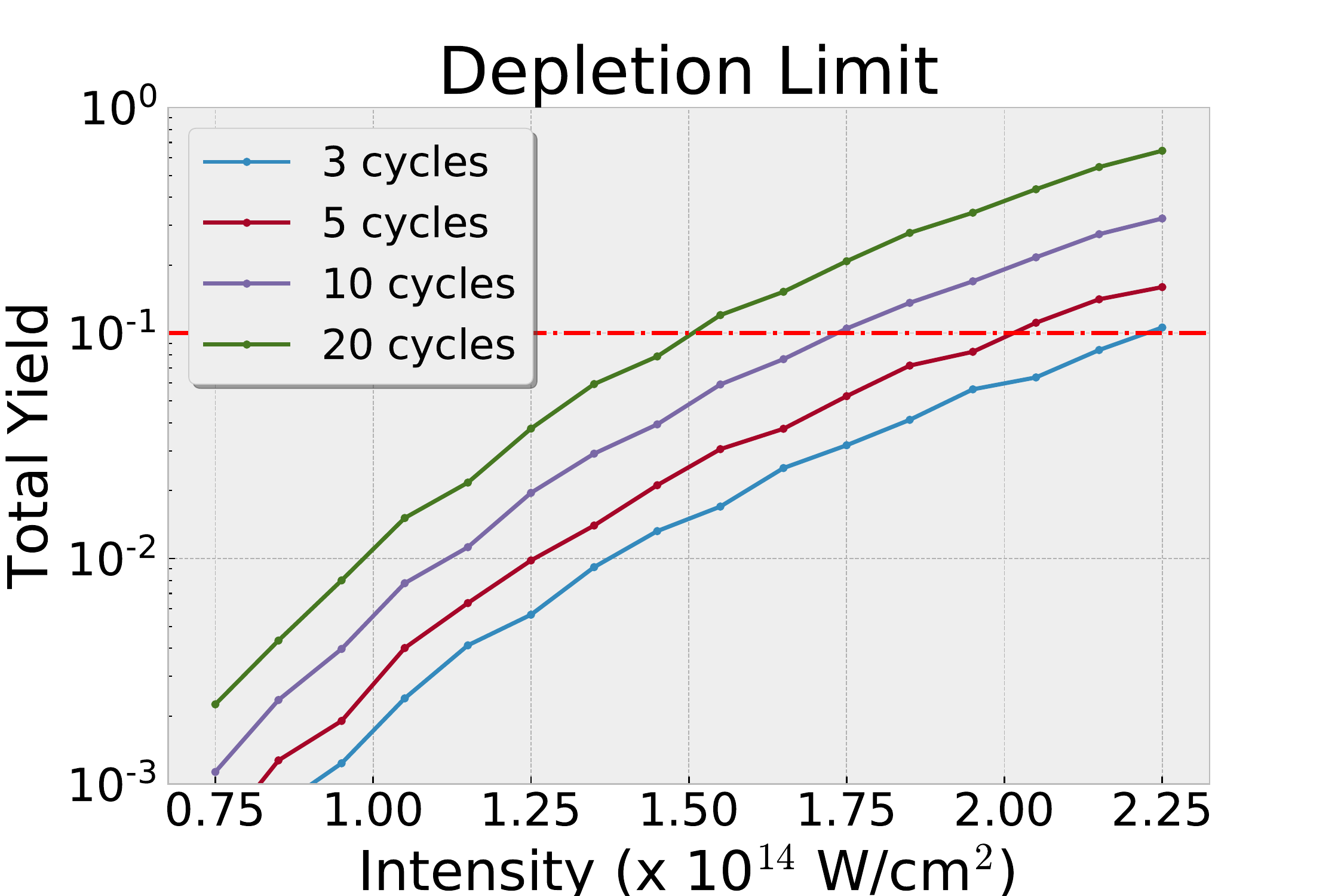}
		\caption{Total ionization yield vs intensity for different length Gaussian pulses, in figure legend. This is computed by integrating the momentum probability distribution. A log scale has been used for the total yield. The red line marks 10\% total ionization. For any laser parameters that lead to a yield above the Fisher information computations will start to break down. The remaining field and target parameters are the same as \figref{fig:Rate_Trends}.}
		\label{fig:depletion}
	\end{figure}
	In \figref{fig:fullupincrease0} we look at the relative uncertainty in the laser intensity as the intensity itself varies. The main reason for considering laser intensity as a measurement variable is the high sensitivity of non-linear ionization processes on the laser intensity. Thus, it should be expected that there is a very strong dependence of the laser intensity uncertainty when varying the intensity. Indeed, we find in the region of interest the quantum Fisher information roughly follows a $I^{-7/2}$ power scaling.
	The $I^{-7/2}$ power scaling will not hold over all intensity ranges but it allows us to demonstrate the high sensitivity of the laser intensity uncertainty.  	
	Given we are altering the parameter that is being measured and computing relative uncertainty, the error is divided by $\up$ which will show up in this scaling.	

	\begin{table*}
		\begin{tabular}{cccccccccc}
			\toprule
			I($10^{14}$~W/cm$^2$) &  Exp (\%) &    QF (\%) &  IF full (\%) &  IF 0.1 (\%) &  IF yield (\%) &  IF spec full (\%) &  IF spec coarse (\%) \\
			\hline
			1.13 &     1.77 &  0.040 &     0.20 &    0.38 &     0.58 &   0.31 &   0.51 \\
			1.53 &     1.31 &  0.014 &     0.062 &    0.11 &     0.20 &   0.091 &   0.11 \\
			1.82 &     1.10 &  0.011 &     0.048 &    0.074 &     0.19 &   0.068 &   0.10 \\
			\hline
		\end{tabular}
		\caption{The relative uncertainty experimentally determined in \cite{Pullen2013} and calculated using the quantum Fisher information (QF) and the full (IF full), coarse (IF $0.1$, where the resolution is denoted in atomic units), yield (IF yield), full spectra (IF spec) and coarse spectra (IF spec coarse, the resolution is set by the experiment and varies between each intensity data set) classical Fisher information. The intensities are given in ($10^{14}$~W/cm$^2$ units)  and cover the lowest three values in the range covered by \cite{Pullen2013}. The number of measurements is set as $4.3\times 10^5$, $4.1\times 10^5$ and $2.1\times10^5$, from top to bottom, corresponding to the number of measurements in each data set.}
		\label{tab:PullenComparison}
	\end{table*}
	
	Beyond a certain laser intensity and pulse length depletion of the target will start to play a role and the trend of reducing uncertainty will not persist in either case. 
	This will act to limit the quantum and classical Fisher informations beyond a particular combination of laser intensity and pulse length.
	Given our model does not account for depletion we must ensure the total ionized portion of the wavefunction stays well below 1.	
	In \figref{fig:depletion} we show the total probability yield of the ionized portion of the wavefunction. A line at 10\% ionization yield is plotted on the figure, if the yield is above this line then the Fisher information computations should be treated with care.
	In \figref{fig:depletion} the 3, 5, 10 and 20 cycle Gaussian pulses reach 10\% ionization at $2.25\times10^{14}$, $2.0\times10^{14}$, $1.75\times10^{14}$ and $1.5\times10^{14}$ ~W/cm$^2$, respectively. This gives a limit on what can be realistically predicted with this model. Given these limitations it is possible to project the lowest possible uncertainty theoretically achievable, within the bounds of this model. For a 5-cycle pulse at $2.0\times10^{14}$~W/cm$^2$ given $5\times 10^4$ measurements the relative uncertainty derived from the quantum Fisher informations would be $2.8 \times 10^{-3}$\%, while for the full momentum measurement it is $7.4\times10^{-3}$\%.	
	It should be noted that assumptions are made and some experimental considerations neglected to obtain this very low uncertainty. For example, it is assumed that the theory is a very good match can be fitted to the experiment to obtain the uncertainty. However, there will be mismatch and it is very difficult to account for how this will affect the final uncertainty.
	This will depend a lot of the type of measurement. The presence of the residual Coulomb potential will increase the ionization probability due to a different barrier and possible resonances with highly excited states, while leading to additional interfering pathways. Some of these pathways will be rescattered, which will further raise the probability at higher photoelectrons energy. On the other hand, the inter-cycle ATI rings will remain invariant \cite{Lai2015,Maxwell2017,Faria2020}.
	Overall such changes would seem to actually further increase the Fisher information (decreasing the uncertainty) given the increased ionization probability as well as additional interferences and `classical' boundaries encoding the ponderomotive energy. A related effect can be seen in the comparison between \tabref{Tab:Uncertainty_Rate} and \tabref{Tab:Uncertainty_ATI}, where there are half the number of ionization channels in the former as in the latter, which leads to radically different moment distributions and reductions in the uncertainty by up to a factor of two. 

In addition to the model, a significant limitation in predicting uncertainties is pulse-to-pulse intensity fluctuations, in particular longer term systematic drift can be catastrophic to the uncertainty. In the review \cite{Kielpinski2014} this was quantified for their specific experimental setup using the Allen deviation \cite{Allan1966} of the laser intensity uncertainty, which demonstrates that the uncertainty stops following the shot noise limit (inversely proportional to the square root of the number of measurements \eqref{eq:cramer-rao}) after around $40$~s of acquisition time, while after $60-100$~s of acquisition time the minimum value is reached and the uncertainty actually starts to increase. This means experiments must limit the acquisition time and therefore counts for accurate intensity measurement, thus optimization of the Fisher information becomes even more important for decreasing the uncertainty.

\subsection{Replicate previous work/ predictions}
\label{sec:NumericalResults:Predicitons}

In this section we will compare the relative laser intensity uncertainty from our computations with that obtained in \cite{Pullen2013}.
It is important to note that we do not expect precise agreement due to the above stated approximations in our methods. This comparison is intended to give an example of how the uncertainty framework can be used and is focused on the difference in general trends.
In \cite{Pullen2013} a TDSE generated spectrum was fitted to experimental results to compute the uncertainty for intensities in the range $1\times 10^{14}$ -- $5\times 10^{14}$ W/cm$^2$ for a 6.3~fs (2--3 cycle) pulse. For each data point in the spectra $10^4$ measurements were made, there are between 6 -- 46 points in each spectra, the number of measurements in the Cram\'ers-Rao bound \eqref{eq:cramer-rao} is set accordingly. The error quoted is $1\%$ over the whole range of intensities. However, a closer look does reveal a slight trend of higher accuracy for higher intensity, as predicted in the previous section. In \tabref{tab:PullenComparison} we compute the uncertainty using the quantum and classical Fisher information for intensities in the range quoted and a 3-cycle pulse length near the $6.3$ fs given in the experiment.

In \tabref{tab:PullenComparison} the uncertainties achieved in \cite{Pullen2013} are listed in the column \textit{Exp (\%)}. The final column \textit{IF spec coarse} is constructed to mimic the experimental measurement procedure. For these values a coarse spectrum measurement was used for the classical Fisher information, where the measurement bin was set to the same size as used in each intensity data set. The values are in all cases below that of the experiment. 
For the lowest intensity the uncertainty is over 3 times below that of experiment and this difference increases with the intensity.
As previously stated, care must be taken given we are idealizing the experimental set up and the ability to match to theory. In particular,
there are a range of incoherent averaging effects, which must be account for. This includes pulse-to-pulse intensity fluctuations, CEP variations, and change of intensity over the focal volume.
The most serious of this is the longer term intensity fluctuation laser drift, which we already mentioned.  The experiment \cite{Pullen2013} did limit acquisition time, however it is likely there is still some deviation from the shot noise limit that contributes to the difference between the experiment and theory.

Focal averaging, CEP variation and short term intensity fluctuations can all be taken into account via incoherent ensembles. In the appendix we extend the equation for the classical Fisher information to show how these may be accounted for in the uncertainty.
In the experiment \cite{Pullen2013} it was necessary to explicitly account for the focal volume of the laser in this experiment to achieve these uncertainties \cite{Pullen2013}. This is the incoherent combination of photoelectron signal from regions of different intensities in the laser focal volume, which leads to a mixed state.
This effect increases with the laser intensity \cite{Kopold2002}, which is consistent with the less good match in the uncertainty for larger laser intensity. This along with depletion explains why the experiment does not follow the trend for lower uncertainties at high intensity. It is possible to design experiments that minimize focal averaging, which could lead to further reductions in the laser intensity uncertainty.

On the other hand, in \cite{Pullen2013} CEP stabilized measurement were used to predict that CEP fluctuations contribute around $10\%$ to the uncertainty for an individual measurement leading to a $\approx0.1\%$ contribution to the overall uncertainty. So it is doubtful this is the cause of the difference between experiment and theory. Similarly, background counts were considered to have a vanishingly small effect in \cite{Kielpinski2014}. Thus, the primary causes of error leading to the difference between experiment and theory will be focal averaging, intensity fluctuations (long and short) and depletion. 

The uncertainty from classical Fisher information measurements in \tabref{tab:PullenComparison} follow the expected pattern, with the yield being the highest, followed by the coarse spectral (IF spec $0.05$) and then the coarse momentum measurement (IF $0.1$), full spectral and finally the full momentum measurement giving the most accuracy. 
It is apparent from \tabref{tab:PullenComparison} that it is also possible to achieve considerable improvement in the uncertainty if another measurement basis is used. The quantum Fisher information reducing to around one hundredth of a percent for the highest intensity. The large difference between QF and IF is likely due to the short pulse length, exhibiting less intercycle interference. Thus, another way to decrease the uncertainty in these measurements would be to use longer pulse lengths.

\section{Conclusions}
\label{sec:Conclusions}
In this work we have developed a new framework for evaluating and understanding the uncertainty of \textit{in situ} measurements in attoscience. The framework we have presented uses the tools and language of quantum sensing, common in quantum optics and quantum information and associated with high precision measurements \cite{Giovannetti2011,Degen2017,Bongs2019,Demkowicz2015, Braun2018, Pezze2018, Pirandola2018}. Recently, as attoscience has matured, both experiment and theory have become more precise and the conversation has turned from qualitative to quantitative measurements. Thus, it is timely and appropriate to consider a framework for understanding and controlling uncertainty in strong-field \textit{in situ} measurements. This is particularly true for laser intensity uncertainty, where many improvements can be made and a unified procedure of determining laser intensity via \textit{in situ} measurements would be very desirable.

Using this framework we explored many illuminating trends regarding the physical processes that govern the uncertainty in \textit{in situ} measurements.
Primarily, we demonstrated how interference of the photoelectron pathways can have a dramatic effect on the laser intensity uncertainty, as information about the latter builds up in the relative coherent phases of the electronic state.
In particular, these phases arise due to the coherent superposition between the bound and continuum states, which allow the continuum states acquire a phase difference we refer to as the ponderomotive phase.
Interference is one of the most commonly exploited ways for determining quantum observables and it is no exception in strong field physics. It is already well-known that intercycle interference/ ATI rings have strong laser intensity dependence \cite{Lewenstein1995}. Here, we make quantitative statements showing this interference vastly reduces the laser intensity uncertainty, by over 20 times in some examples. We also show subcycle or intracycle interference leads to further reductions in the uncertainty. 

Due to the great flexibility of this framework we are able to investigate many types of measurement, in particular we accounted for uncertainty in the measurement apparatus via coarse measurements. Thus, we can make direct statements about how the uncertainty in the measurement equipment affects the final laser intensity uncertainty. We found that beyond a certain cut-off additional precision will not strongly affect the uncertainty. This is a stepping stone from which it would be possible to use this framework to explicitly prototype detectors and optimize their efficacy for determining particular parameters. 
Comparisons of resolutions is a prime example of how this methodology can be used to evaluate different types of measurement/ detection for parameter estimation.
By modelling the changing resolution present in detectors \cite{Wolter2015}, experimental setups could be optimized for particular measurement.
We foresee that this type of method could even be used in prototyping to guide the best detector designs given different measurement goals.
In another optimization example, this framework could be used to determine whether a trade-off between increased control or precision in the measurement vs a reduction in counts is worth it in terms of how it affects the uncertainty of a measurement variable.

We also evaluated trends in the laser intensity uncertainty over increasing pulse length and laser intensity, and we find increasing sensitivity as both these parameters increase. 
These trends are expected and unsurprising and as such are a good evaluation of the validity of the approach.
Furthermore, direct comparison was made to \textit{in situ} measurements made in \cite{Pullen2013}, where the computation of the uncertainty was designed to follow the measurements made therein. Reduction in the uncertainty by 2-3 times may be possible by using high resolution 3D momentum measurements, while measuring in another non-momentum basis could give the capacity for a further 4-5 times reduction. Choosing a longer pulse length would yet further reduce the uncertainty. 

It is important to note that due to effects like depletion, pulse-to-pulse fluctuations and intensity variation over the focal volume, some care should be taken when comparing our theoretical predictions with experiment.
We assume we have access to multiple copies of an identical wave function, when in fact, there are incoherent averaging effects leading to a mixed state. Quantum metrology and the Fisher information with mixed states has been studied in \cite{Fiderer2019, Modi2011}.
Laser drift can have a significant effect on the intensity uncertainty \cite{Kielpinski2014}, which may be mitigated by limiting the acquisition time. However, this limits the number of counts meaning it becomes very important to optimize the uncertainty via the Fisher information. On the other hand, as laser systems progress to higher power stability and higher repetition rates (see e.g. \cite{Wolter2015, Baudisch2015,Ciriolo2017, Mero2018}) the number of counts can be increased.
Focal volume effects can be minimized in the experiment by reducing the size of the interaction region between the laser and gas jet \cite{Kielpinski2014}. Alternatively, one could shape the laser intensity spatial profile to exhibit a flat top (see \cite{Boutu2011,Toma1999} for examples). To account for CEP averaging in experiment one may stabilize the CEP for a single value (although this will reduce the counts and ultimately may not be worth doing).
The effects of noise/ background counts and detector efficiency was not accounted for in this first approach, which will increase the computed uncertainties. A thorough review of many of the possible sources of noise as well accounting for detector sensitivity and signal-to-noise ratio can be found in \cite{Degen2017}.

Additionally, the use of the SFA leaves out effects such as Coulomb phases and distortions of the photoelectron, coupling with bound states and multi-electron interaction. The former maybe be addressed by using Coulomb-corrected approaches such as the CQSFA \cite{Lai2015, Lai2017,Maxwell2017,Maxwell2017a, Maxwell2018, Faria2020} or directly employing numerical solutions of the time-dependent Shr\"odinger equation such as with the Qprop software \cite{BauerQprop2006}. However, computations of the Fisher information the integrals are numerically intensive even for the simple SFA model. Thus, faster quantum orbit models such as the CQSFA would be more appropriate and feasibility studies should be performed before attempting this with a numerical TDSE solver.
Future work should focus on tackling the above-stated issues, in addition to implementing more detailed measurement procedures to better reflect experiment. This would enable quantitative modelling of uncertainty \textit{in situ} measurement allowing for high levels of optimization.

Nonetheless, the framework for evaluating uncertainties presented here reveals the key physical processes involved for an ideal strong field system. The ponderomotive phase the photoelectron acquired in the continuum lead to the sensitive quadratic scaling with time. Beyond laser intensity, this opens the door for evaluating any parameters of interest in strong field ionization. Taking this one step further multiple field or target parameters could be measured, which is often desirable in attoscience. This theoretical means of calculating the uncertainty, will go side-by-side with the fitting procedures involved \textit{in situ} measurements, allowing the derived uncertainties to be understood, giving the potential for them to be considerably improved.

\begin{acknowledgments}
We would like to thank Marco Genoni for bringing some of the latest literature to our attention. We would like to give our utmost thanks to Maciej Lewenstein for very fruitful discussion and his help with the publication process.
This work was in part funded by the UK Engineering
and Physical Sciences Research Council (EPSRC).  
ASM acknowledges grant EP/P510270/1, which is within the remit of the InQuBATE Skills Hub for Quantum Systems Engineering. CFMF would like to acknowledge EPSRC grant EP/T019530/1. SB would like to acknowledge EPSRC Grants No. EP/N031105/1 and EP/S000267/1.

ASM as part of the ICFO group acknowledges support from ERC AdG NOQIA, Agencia Estatal de Investigación (``Severo Ochoa'' Center of Excellence CEX\allowbreak{}2019-000910-S, Plan National FIDEUA PID2019-106901\allowbreak{}GB-I00/10.13039 / 501100011\allowbreak{}033, FPI), Fundació Privada Cellex, Fundació Mir-Puig, and from Generalitat de Catalunya (AGAUR Grant No.\ 2017 SGR 1341, CERCA program, QuantumCAT \_U16-011424, co-funded by ERDF Operational Program of Catalonia 2014-2020), MINECO-EU QUANTERA MAQS (funded by State Research Agency (AEI) PCI2019-111828-2 / 10.\allowbreak{}13039/\allowbreak{}501100011033), EU Horizon 2020 FET-OPEN OPTOLogic (Grant No 899794), and the National Science Centre, Poland-Symfonia Grant No.\ 2016/20/W/ST4/00314. 

\end{acknowledgments}

\appendix
\section{Incoherent Effects}
In this appendix we consider macroscopic incoherences which will affect the final derived uncertainty and suggest how they can be accounted for in the Classical Fisher information computations.
\subsection{Focal Averaging}
The intensity will vary over the focal volume, thus a correct treatment will consider incoherent emission from across the focal volume \cite{Kopold2002,Kielpinski2014}. This will of course affect the computation of the classical Fisher information. The probability for ionization over the whole focal volume can be written as an integral over the probability given a specific laser intensity
\begin{equation}
	\mathcal{P}_{FA}(\mu|I_0) = \int \dd^3 \rb n(\rb) \mathcal{P}(\mu | I(\rb, I_0)),
\end{equation}
where $n(\rb)$ is the atom number density and $I(\rb)$ is the variation of the laser intensity over the focal volume, this may be approximated by a Gaussian beam profile or alternatively by a profile measured in experiment. A Gaussian beam profile is given by
\begin{equation}
	I(\rho, z, I_0) = I_0\left(\frac{w_0}{w(z)}\right)
						\exp\left(-\frac{2\rho^2}{w(z)^2}\right),
\end{equation}
written in cylindrical spacial coordinates where the beam waist is given by
\begin{equation}
	w(z)=w_0(1+(z/z_0)^{1/2}),
\end{equation}
$w_0=\sqrt{\lambda z_0/\pi}$ and $z_0$ denotes the Rayleigh range. It is common the approximate the number density as constant in the interaction region, which may be reasonable for atomic beams or gas cells.

\subsection{CEP Averaging}
For short pulses the carrier-envelope phase (CEP) strongly affects the final probability distributions, and it will be averaged over in experiments unless it is stabilized. The probability distribution can be integrated over all values to account for this
\begin{equation}
	\mathcal{P}_{CEP}(\mu|I) = \int_0^{2\pi} \dd \phi \mathcal{P}(\mu | I, \phi).
\end{equation}

\subsection{Intensity Fluctuations}
Long terms drift of the laser intensity can be catastrophic for uncertainty and we discuss its implications in \secref{sec:NumericalResults:GeneralTrends}. Short term fluctuations are less of an issue and can be approximately accounted for in the following way. For particular experiments the root-mean-square error (RMSE), denoted $\sigma$ here may be known. This can be used to approximate the deviations by a distribution $f(I|I_0, \sigma)$ and incorporated into the probability
\begin{equation}
	\mathcal{P}_{IF}(\mu|I_0) = \int_{I_0-\Delta}^{I_0+\Delta} \dd I f(I|I_0, \sigma) \mathcal{P}(\mu | I),
\end{equation}
where $\Delta$ is fixed to capture the vast majority of intensity fluctuations.
To first approximation the distribution may be assumed to be Gaussian
\begin{equation}
	f(I|I_0, \sigma) = \frac{1}{\sigma\sqrt{2\pi}}
	 e^{-\frac{1}{2}\left(\frac{I-I_0}{\sigma}\right)^2}.
\end{equation}
For modern laser systems \cite{Baudisch2015,Ciriolo2017, Mero2018} the RMSE of the laser power is typically under $1\%$ for some hours of acquisition time. However, variations in other experimental parameters (e.g. beam waist, pulse length, temperature) can cause additional fluctuations of the laser intensity in the interaction region.

\subsection{Incorporating in to the Classical Fisher Information}
All these incoherent effects can be straight forwardly combined
\begin{align}
	&\mathcal{P}_{All}(\mu|I_0) =\notag\\
	&\int_{I_0-\Delta}^{I_0+\Delta} \dd I' f(I'|I_0, \sigma)
	\int \dd^3 \rb n(\rb) 
	\int_0^{2\pi} \dd \phi
	\mathcal{P}(\mu | I(\rb, I'), \phi),
	\intertext{which we will write more simply as}
	&\mathcal{P}_{All}(\mu|I_0)=\int_{I_0-\Delta}^{I_0+\Delta} \dd I' f(I'|I_0, \sigma) \mathcal{P}_0(\mu | I'),
\end{align}
where $\mathcal{P}_0(\mu | I')$ contains the averaging over the focal volume and the CEP.
In order to compute the Fisher information we must take the derivative of the probability with respect to $I_0$
\begin{align}
	\frac{\partial \mathcal{P}_{All}(\mu|I_0)}{\partial I_0} &= \int_{I_0-\Delta}^{I_0+\Delta} \dd I' 
	\frac{\partial f(I'|I_0, \sigma)}{\partial I_0} \mathcal{P}_0(\mu | I')\notag\\
	&+\left.f(I'|I_0, \sigma) \mathcal{P}_0(\mu | I')\right|_{I_0-\Delta}^{I_0+\Delta}.
\end{align}
The terms evaluated at the boundaries can be removed if $\Delta$ is big enough. Then $\mathcal{P}_{All}(\mu|I_0)$ and $\partial_{I_0}\mathcal{P}_{All}(\mu|I_0)$ can be straight forwardly plugged into \eqref{eq:ClassicalFisher}.


\bibliography{MesLasIntensity}
\end{document}